\documentclass{ws-rv975x65}
\usepackage{subfigure}     
\usepackage{ws-rv-van}     

\usepackage{bibentry}
\usepackage{bm}
\usepackage{color}
\usepackage{hyperref}

\definecolor{clear}{gray}{1}

\renewcommand{\case}[2]{\ensuremath{{\textstyle\frac{#1}{#2}}}}
\newcommand{\eighth}{\ensuremath{\case{1}{8}}}
\newcommand{\half}{\ensuremath{\case{1}{2}}}
\newcommand{\Det}{\ensuremath{\mathop{\mathrm{Det}}}}
\newcommand{\tr}{\ensuremath{\mathop{\mathrm{tr}}}}

\newcommand{\onlinecite}[1]{\nocite{#1}\citenum{#1}} 

\newcommand{\kronfeldtitle}{Lattice Gauge Theory and the Origin of Mass}

\begin{document}


\chapter[Lattice Gauge Theory]{\vspace*{-0.2in}\kronfeldtitle\footnotemark}   
\label{kronfeld}
\footnotetext{Prepared for \emph{One Hundred Years of Subatomic Physics}, edited by Ernest Henley and 
Stephen Ellis.}
\vspace*{-1.2in}
\fcolorbox{clear}{clear}{\LARGE\color{clear} Chapter 1 \quad Chapter 1 \; Chapter 1}
\vspace*{0.5in}

\author[A. S. Kronfeld]{Andreas S. Kronfeld}

\address{Theoretical Physics Department, Fermi National Accelerator Laboratory, \\
Batavia, Illinois, USA}

\begin{abstract}
    Most of the mass of everyday objects resides in atomic nuclei; the total of the electrons' mass adds up 
    to less than one part in a thousand.
    The nuclei are composed of nucleons---protons and neutrons---whose nuclear binding energy, though 
    tremendous on a human scale, is small compared to their rest energy.
    The nucleons are, in turn, composites of massless gluons and nearly massless quarks.
    It is the energy of these confined objects, via $M=E/c^2$, that is responsible for everyday mass.
    This article discusses the physics of this mechanism and the role of lattice gauge theory in 
    establishing its connection to quantum chromodynamics.
\end{abstract}

\body
\section{Introduction}
\label{lft:sec:intro}

With the recent observation of a Higgs-like particle~\cite{Aad:2012qt,Chatrchyan:2012qt,Aaltonen:2012qt},
people from all walks of life are talking about the origin of mass.
Careful accounts note that this new object's underlying field generates mass neither for luminous matter nor
for dark matter but for standard-model particles.
Among these, the top quark and the $W$ and $Z$ bosons are especially intriguing, the storyline goes, because
their masses are similar to those of whole atoms of gold or silver.
But where does the mass of a gold ring or a silver spoon come from? This article reviews our understanding of
the origin of mass of these and all other everyday objects, starting from first principles.

The density of gold metal is around $20$~g\,cm$^{-3}$.
At the beginning of the twentieth century, no one knew how mass is distributed within atoms, and several
ideas had been put forth~\cite{Thomson:1904pp,Nagaoka:1904sr}.
Then an experiment carried out by Hans Geiger and Ernest Marsden found an astonishing rate of wide-angle
scattering of a beam $\alpha$ particles incident on a gold foil~\cite{Geiger:1909ab}.
Their laboratory director, New Zealander Ernest Rutherford, realized that their findings could be understood
if atoms contain a massive nucleus surrounded by a cloud of electrons~\cite{Rutherford:1911zz}.
The density of nuclear material ranges from $20\times10^{13}$~g\,cm$^{-3}$ for a gold nucleus to
$60\times10^{13}$~g\,cm$^{-3}$ for an isolated proton.
One cubic millimeter (the size of a coarse grain of sand) of such nuclear material weighs about as much as
two aircraft carriers.

The discovery of the neutron~\cite{Chadwick:1932ma} showed that atomic nuclei consist of protons and
neutrons, bound together by the so-called strong force.
The forces in the nucleus generate tremendous energy, yet nuclear fission releases only around one part in a
thousand of the total rest energy.
Nuclear fusion producing ${}^4$He releases a larger fraction of the total nuclear rest energy, but still less
than~1\%.
Thus, the origin of the bulk of nuclear mass lies beyond nuclear chemistry and more deeply within the
nucleons themselves.


Of course, the nucleon has structure too.
Indeed, deeply inelastic electron-nucleon scattering measurements (wide angles again) are modeled well with
weakly interacting constituents known as partons~\cite{Feynman:1969ej,Feynman:1972pm}.
To obtain a full appreciation of the interior of the nucleon, however, one must to turn to the modern theory
of the strong interactions, namely quantum chromodynamics (QCD).
QCD merges the ideas of the quark model (introduced to account for the plethora of strongly-interacting
hadrons~\cite{Zweig:1981pd,Zweig:1964jf,Gell-Mann:1964nj}), the quantum number ``color'' (introduced to
reconcile spin and statistics~\cite{Greenberg:1964pe,Han:1965pf}), and partons into a self-contained
theory~\cite{Bjorken:1969ja}.

The Lagrangian of QCD~\cite{Fritzsch:1973pi} looks like that of quantum electrodynamics (QED).
In both cases, the interactions are specified by a gauge symmetry, SU(3) for QCD and U(1) for QED.
SU($N$) is the nonabelian group of $N\times N$ unitary matrices with unit determinant.
As a consequence of the nonabelian, i.e., noncommuting, nature of SU($N$), the quanta of the gauge
field---known in QCD as gluons---carry color~\cite{Yang:1954ek,RonShaw}.
Because QED's U(1) group of phase factors commutes, the gauge quantum is electrically neutral, in accord with
the natural behavior of the photon.
The self-coupling of the gluon is responsible for the markedly different dynamics in QCD.
In particular, quantum effects, which can be examined in one-loop perturbation theory, render the QCD
coupling smaller and smaller at short distances~\cite{Politzer:1973fx,Gross:1973id}.
This ``asymptotic freedom'' means that QCD reproduces the simplicity of the parton model.

The flip side of asymptotic freedom is that the strong interaction strengthens at large distances.
A~``typical mass scale $\Lambda_\mathrm{QCD}$'' separates weak from strong coupling.
At distances large enough so that the coupling is strong, perturbative techniques are insufficient to
understand fully what happens.
Nevertheless, the strengthening of the force provides a hint that it is possible to explain not only the
origin of hadronic mass but also why isolated quarks are never observed (known as confinement).
This article discusses how, a century after the Geiger-Marsden experiment, we have established a
connection from the QCD Lagrangian to the mass of the atomic nucleus and, hence, all everyday objects.
Indeed, this connection sheds light on confinement as well.
The central theoretical and conceptual tool is lattice gauge theory~\cite{Wilson:1974sk}, which enables
nonperturbative calculations via a mathematically rigorous definition of quantum field theory.
The calculations lie beyond the scope of pencil and paper and, in fact, rely on leadership-class
supercomputers.

The rest of this article is organized into two main parts.
Section~\ref{lft:sec:lattice} recalls the early (and prehistoric) development of lattice gauge theory.
Section~\ref{lft:sec:qcd} reviews QCD calculations based on lattice gauge theory, with special attention to 
calculations the shed light on the origin of (everyday) mass.
The Appendix recounts a tale about lattice field theory, Werner Heisenberg, and a children's puzzle.

\section{Lattice Gauge Theory}
\label{lft:sec:lattice}

Before turning to lattice gauge theory itself, it is helpful to discuss asymptotic freedom a bit more.
Let us start with the relation between the bare gauge coupling and a renormalized coupling.
A physical renormalization scheme comes from the force $F(r)$ between static source and sink of color,
separated by a distance~$r$,
\begin{equation}
    r^2 F(r) = -\frac{C_F}{4\pi}\,g_F^2(r),
\end{equation}
where $C_F=1$ for U(1), $C_F=(N^2-1)/2N$ for SU($N$).
In perturbation theory, the force arises from one-gluon exchange and from Feynman diagrams with loops.
To define the loop integrals, one must have an ultraviolet cutoff.
A~lattice with spacing~$a$ builds one from the outset.
Adopting lattice notation, the relation between $g_F^2$ and the bare coupling $g_0^2$ can be written as
follows:
\begin{equation}
    g_F^{-2}(r) = g_0^{-2}(a) + \beta_0 \ln(a^2/r^2) + c_{F\leftarrow0} + \mathrm{O}(g^2).
    \label{lft:eq:Ffrom0}
\end{equation}
The constants $\beta_0$ and $c_{F\leftarrow0}$ stem from the one-loop diagrams, and the omitted terms from
diagrams with two or more loops.
For what follows, $c_{F\leftarrow0}$ is not very important, but the sign of $\beta_0$ is key.
Direct calculation in SU($N$) gauge theories yields~\cite{Politzer:1973fx,Gross:1973id} 
\begin{equation}
    \beta_0 = - \frac{2}{3} \frac{n_f}{16\pi^2} + \frac{11}{3}\frac{N}{16\pi^2},
    \label{lft:eq:beta0}
\end{equation}
where $n_f$ is the number of quark flavors.
In QED, the second term, which stems from the gluon loop, is absent, and $n_f$ is replaced with
$2\sum_lq_l^2$, where $q_l$ is the electric charge of charged particle $l$ (e.g., the electron $q_e=-1$).
In QCD with $n_f\le16$, one finds $\beta_0>0$, which yields asymptotic freedom, namely $g_F^2(r)$ decreases
as $r$~decreases.
In QED (and in QCD with $n_f>16$), $\beta_0<0$.

Renormalization of the bare gauge coupling $g_0^2(a)$ makes the right-hand side of 
Eq.~(\ref{lft:eq:Ffrom0}) independent of the lattice spacing $a$.
Then one can write
\begin{equation}
    g_F^{-2}(r) = \beta_0 \ln(r^{-2}/\Lambda_F^2),
\end{equation}
where a scale $\Lambda_F$ appears
\begin{equation}
    \Lambda_F = a^{-1} \, e^{-1/2\beta_0g_0^2(a)} e^{-c_{F\leftarrow0}^{}/2\beta_0}.
    \label{lft:eq:LambdaF}
\end{equation}
If $\beta_0<0$ as in QED, this scale is commensurate with the ultraviolet cutoff~\cite{GellMann:1954fq}.
On the other hand, if $\beta_0>0$, as in QCD with the six observed quark flavors, the scale~$\Lambda_F$ is
much smaller than the cutoff.
Such hierarchies of scale are an essential feature of renormalization in a more general, nonperturbative
context~\cite{Wilson:1969zs,Wilson:1970ag}.

Different renormalization schemes lead to different scales~\cite{Celmaster:1979km}.
In a scheme ``$R$'',
\begin{equation}
    \Lambda_{R} = \Lambda_F \, e^{-c_{R\leftarrow F}^{}/2\beta_0},
\end{equation}
where $c_{R\leftarrow F}=c_{R\leftarrow0}-c_{F\leftarrow0}$ is regulator independent.
For small exponents, such scales are quantitatively similar.
Qualititatively, the range of such scales marks the transition from weak to strong coupling and is
usually called~$\Lambda_\mathrm{QCD}$.

The discovery of asymptotic freedom spawned widespread interest in all aspects of QCD, including applications
to high-energy scattering processes~\cite{Ellis:1991qj} as well as puzzles such as the nonobservation of
isolated quarks~\cite{Perl:2009zz}.
Kenneth Wilson, who had been working on critical phenomena, was one of those who (re)directed his attention
to the strong interactions.
When reading his 1974 paper introducing lattice gauge theory~\cite{Wilson:1974sk}, it may look as though he
developed lattice gauge theory to study confinement.
In 2004, however, he reminisced~\cite{Wilson:2004de}

\begin{quote}
    The discovery of asymptotic freedom, made possible by earlier developments on the renormalizability of 
    non-Abelian gauge theories by Veltman and 't~Hooft,$^[$\cite{'tHooft:1971fh,'tHooft:1972fi}$^]$ made it 
    immediately clear, to me as well as many others, that the preferred theory of strong interactions was 
    [QCD]. \ldots 

    Unfortunately, I found myself lacking the detailed knowledge and skills required to conduct research 
    using renormalized non-Abelian gauge theories.
    My research prior to 1973 had not required this knowledge so I had never spent the time necessary to 
    acquire it.
    
    What was I to do, especially as I was eager to jump into this research with as little delay as possible? 
    I realized that from my prior work in statistical mechanics$^[$\footnotemark$^]$ I knew a lot about 
    working with lattice theories, including the construction of high temperature expansions for such 
    theories.
    I decided I might find it easier to work with a lattice version of QCD than with the existing continuum 
    formulation of this theory. 
    Moreover, this meant I could be doing original research immediately, rather than having to spend weeks 
    or months absorbing other people's research.
\end{quote}
In gauge theories, the ``high-temperature expansion'' of statistical mechanics develops a strong-coupling 
series in powers of~$1/g_0^2$.
\footnotetext{Wilson's work in statistical mechanics started out as an application of his
renormalization-group ideas from particle physics~\cite{Wilson:1969zs,Wilson:1970ag} to critical
phenomena~\cite{Wilson:1971bg,Wilson:1971dh}.
It was very successful, leading to a renormalization-group solution of the Kondo problem of magnetic
impurities in nonmagnetic metals~\cite{Wilson:1974mb} that earned him the 1982 Nobel Prize in
Physics~\cite{Wilson:1993dy}.}

Wilson's 1974 paper~\cite{Wilson:1974sk} showed how to preserve local gauge invariance when spacetime is 
replaced with a lattice. 
The main mathematical ingredient is straightforward.
Matter fields transform under local gauge transformations as
\begin{equation}
    \phi(x) \mapsto g(x)\phi(x),
    \label{lft:eq:gphi}
\end{equation}
where $g(x)$ is an element of a Lie group~$G$, e.g., U(1) or SU($N$).
It is not hard to show that the so-called parallel transporter
\begin{equation}
    U_s(x,y) = \mathbb{P}_s \exp \int_x^y dz\cdot A(z)
    \label{lft:eq:U}
\end{equation}
transforms as
\begin{equation}
    U_s(x,y) \mapsto g(x) U_s(x,y) g^{-1}(y).
    \label{lft:eq:gUg}
\end{equation}
Here $A^\mu(x)$ is the gauge potential, taking values in the Lie algebra of~$G$. 
The path-ordering symbol $\mathbb{P}_s$ prescribes the order of matrix factors in the power series of the 
exponential function to lie along the path $s$ from $x$ to $y$.
From Eqs.~(\ref{lft:eq:gphi}) and~(\ref{lft:eq:gUg}), products of the form $\phi^\dagger(x)U(x,y)\phi(y)$ 
clearly are gauge invariant.
On a lattice, any $U(x,y)$ can be built out of one-link parallel transporters $U(x,x')$, where $x$ and $x'$ 
are nearest-neighbor lattice sites.
The dynamical variables of lattice gauge theory are, thus, matter fields on sites and gauge-group variables 
on nearest-neighbor links.
Note that this construction works for a lattice of any geometry~\cite{Christ:1982ck}.

Wilson was not the first to consider lattice gauge theory.
Wilson knew~\cite{Wilson:1993dy} about work on lattice field theories by Gregor Wentzel~\cite{Wentzel:1940gu}
and by Leonard Schiff~\cite{Schiff:1953zza} for the strongly-coupled meson-nucleon system.
He did not know, until later~\cite{Wilson:1984vw}, about the Ising gauge theory of Franz
Wegner~\cite{Wegner:1971qt}, or about the (unpublished) nonabelian lattice gauge constructions of Jan
Smit~\cite{Smit:1972}\nocite{Smit:2002ug} and of Alexander Polyakov~\cite{Polyakov:1975rs}.

Wentzel's and Schiff's lattice field theories can be traced, via their textbooks~\cite{Wentzel:1949,%
Schiff:1968}, back to Werner Heisenberg and Wolfgang Pauli's first paper on the quantization of 
electrodynamics.
To lend rigor to several (then brand new) mathematical manipulations, they introduced a spatial lattice,
writing~\cite{Heisenberg:1929wf}:

\begin{quote}
    \emph{In der Tat kann man den Fall kontinuierlich vieler Freiheitsgrade, wo die Zustandsgr\"o\ss en
    Raumfunktionen sind, stets durch Grenz\"ubergang aus dem Fall endlich vieler Freiheitsgrade 
    gewinnen.}
    
    Indeed, one can always obtain the case of continuously many degrees of freedom, where the state 
    variables are functions of space, through a limit of the case of finitely many degrees or freedom.
    (\emph{Author's translation}.)
\end{quote}
Heisenberg and Pauli thus introduce fields on a discrete set of cells, whose centers form a lattice, and use
the limit of smaller and smaller cells to establish the functional equations of motion from their discrete
counterpart.
They further employ the lattice to derive the Dirac $\delta$ function in equal-time commutators in field
theory from the Kronecker $\delta$ symbol in quantum mechanics for a finite collection of degrees of freedom.
They do not revisit the lattice when considering local gauge symmetry, so they did not need
Eqs.~(\ref{lft:eq:gphi})--(\ref{lft:eq:gUg}).
They also did not discuss electron fields on the lattice---as we shall see below, prudently.
Heisenberg later pondered the lattice as the fundamental structure of space, as discussed in the Appendix.

Despite the antecedents, Wilson's 1974 paper~\cite{Wilson:1974sk} was a watershed for at least three reasons.
First, Wilson discussed very clearly how to understand color confinement via the energy stored between a
source and sink of color separated by larger and larger distances.
Furthermore, he showed that lattice gauge theories confine in the strong-coupling limit (according to his
criterion).
Second, the spacetime lattice provides a mathematically sound definition of the functional integral.
As with canonical quantization~\cite{Heisenberg:1929wf}, the functional integral for a countable set of
degrees of freedom is straightforward to define, and the one for continuously many degrees of freedom is
obtained as a limit.
Third, Wilson's earlier work on the renormalization group, both in critical phenomena and in the strong
interactions, lent credence to the proposal that the lattice formulation of gauge theory could be connected
to the gauge-fixed, perturbative formulation of gauge theory.
Moreover, all these aspects provided a starting point for others to begin research on lattice gauge
theory~\cite{Rebbi:1983}, particularly after an influential series of
papers~\cite{Balian:1974ts,Balian:1974ir,Balian:1974xw} explained the techniques of statistical mechanics in
particle physics language.

Wilson's criterion for color confinement starts with the parallel transporters in Eq.~(\ref{lft:eq:U}).
The interacting quark propagater from $x$ to $y$ consists of a properly weighted average of~$U_s(x,y)$ over
all paths~$s$.
A~meson propagator consists of an average of a similar object,
$U_s(x,y)U_{\bar{s}}(y,x)=U_{s\cup\bar{s}}(x,y)$, where $s$ ($\bar{s}$) is the path taken by the quark
(antiquark).
Such closed loops of parallel transport are now known as Wilson loops, and even open parallel transporters
$U_s(x,y)$ of shortest path are often known as Wilson lines.
Note that the trace, $\tr U_\mathrm{loop}(x,x)$, of a closed loop is gauge invariant.

It is instructive to consider loops for which the spatial coordinates of $x$ and $y$ are the same and set
$|x-y|=t$.
Furthermore, imagine loops for which $s$ and $\bar{s}$ separate a distance $r$ apart (away from the points
where they join).
Let us denote such a Wilson loop $U_{r\times t}$, because the shape is more pertinent than the path names $s$
and~$\bar{s}$.
If the expectation value (when $t\gg r$)
\begin{equation}
    \langle \tr U_{r\times t}\rangle\sim e^{-Vt}
    \label{lft:eq:perimeter}
\end{equation}
with $V$ independent of $r$ for large~$r$, then quark and antiquark lines can easily separate to large~$r$.
Such behavior obtains in electrodynamics and signals the absence of confinement.
On the other hand, if the expectation value 
\begin{equation}
    \langle \tr U_{r\times t}\rangle\sim e^{-\sigma rt},
    \label{lft:eq:area}
\end{equation}
then large separations of quark and antiquark are highly improbable.
Such behavior is known as the ``area law'' and corresponds to confinement~\cite{Wilson:1974sk}.

To compute $\langle \tr U_{r\times t}\rangle$, one employs the functional-integral formulation of field
theory.
In general, the central objects are correlation functions, so let us introduce several composite,
color-singlet fields $\Phi_i$, where the subscript labels both spacetime and internal indices, as well as
distinguishing one composite field from another.
The $n$-point correlation function is given by
\begin{eqnarray}
    \left\langle\Phi_1\cdots\Phi_n\right\rangle & = & \frac{1}{Z} 
        \int \prod_{x,\mu} dU_{x,\mu} \prod_x d\psi_xd\bar{\psi}_x \,
            \Phi_1\cdots\Phi_n \, e^{-S(U,\psi,\bar\psi)}, 
    \label{lft:eq:vev} \\
    Z & = & \int \prod_{x,\mu} dU_{x,\mu} \prod_x d\psi_xd\bar{\psi}_x \, e^{-S(U,\psi,\bar\psi)}.
    \label{lft:eq:Z}
\end{eqnarray}
Once invariant measures have been specified for the SU($N$) variables~\cite{Haar:1933im} $U_{x,\mu}$ and for
the fermion (quark) variables~\cite{Berezin:1966nc} $\psi_x$ and $\bar{\psi}_x$, the right-hand sides of
these equations are well-defined finite integrals.
Mathematicians would say, ``they exist.'' %
In the limit of infinite temporal extent, such a correlation function yields the vacuum-expectation value of
the time-ordered product of the~$\Phi_i$s.
Note, however, that lattice gauge theory is well-defined in Euclidean spacetime.
The Euclidean signature also leads to minus signs, rather than factors of $i$, in several formulae, but it is
not a limitation in principle.

Equations~(\ref{lft:eq:vev}) and~(\ref{lft:eq:Z}) specify a canonical average and partition function in
classical statistical mechanics.
Such systems can exhibit second-order phase transitions~\cite{Wilson:1973jj}, which are characterized by a
large correlation length $\xi\gg a$, where $\xi$ measures the falloff of a correlation function,
$e^{-|x-y|/\xi}$.
Correlation functions with different symmetry-group representations can have different correlation lengths,
but all $\xi_\alpha\gg a$.
At second-order phase transitions, the details of the lattice (e.g., whether the crystal is triclinic,
hexagonal, or cubic) become unimportant and the long-distance behavior depends only on the internal
symmetries of the interactions.
These phenomena translate into particle physics language as follows.
One identifies the inverse correlation lengths~$\xi_\alpha^{-1}$ with particle masses~$M_\alpha$, because a
particle two-point function behaves like $e^{-M_\alpha|x-y|}$.
Thus, to define a continuum quantum field theory via lattice field theory, one seeks a point in the space of
couplings, such that a hierarchy between the particle masses and the inverse lattice spacing emerges.
The hierarchy of scale is the key feature, while Eq.~(\ref{lft:eq:LambdaF}) gives the specific example
relevant to asymptotically free theories.
It means that $M_\alpha\propto\Lambda_F$, with coefficients that depend on the chromodynamics of quarks and
gluons, but not on the lattice.

Because the integrals in Eqs.~(\ref{lft:eq:vev}) and~(\ref{lft:eq:Z}) exist, they provide a platform for a
rigorous construction of quantum field theory~\cite{Seiler:1982pw}.
To reconstruct a quantum-mechanical Hilbert space from a Euclidean field theory, the functional integrals
must enjoy certain properties, such as a positive action, known as the Osterwalder-Schrader
axioms~\cite{Osterwalder:1973dx,Osterwalder:1974tc}.
The simplest lattice actions satisfy these conditions~\cite{Osterwalder:1977pc}.
That said, the challenge is to lend mathematical rigor to the limiting procedure of the \emph{renormalized}
continuum limit, i.e., one requires a rigorous understanding of critical phenomena.
For a historical review of this field, including the role of the renormalization group, see
Ref.~\onlinecite{Summers:2012vr}.

Let us return to Eq.~(\ref{lft:eq:vev}) and examine the average of the Wilson loop to learn whether (lattice)
gauge theory confines.
Taking a hypercubic lattice and a simple gauge-invariant lattice Lagrangian, chosen to reduce to
Yang-Mills Lagrangian~\cite{Yang:1954ek,RonShaw} in the classical continuum limit, Wilson found
\begin{equation}
    \langle \tr U_{m\times n} \rangle = (2N/g_0^2)^{mn} = e^{-\ln(g_0^2/2N)mn}.
    \label{lft:eq:Umn}
\end{equation}
for an $m\times n$ rectangular Wilson loop.
In fact, the same calculation shows that for any planar Wilson loop, the (dimensionless) area replaces $mn$
in Eq.~(\ref{lft:eq:Umn}).

The area law follows from a simple property of invariant integration over the gauge group, which states 
\begin{equation}
    \int dU \,U = 0, 
    \label{lft:eq:haar}
\end{equation}
for example, $\int_{-\pi}^{\pi}d\theta\,e^{i\theta}=0$ for U(1), and
$\int_{-\pi}^{\pi}d\theta\int d^2\hat{n}\,e^{i\bm{\hat{n}}\cdot\bm{\sigma}\theta}=0$ for SU(2).
Equation~(\ref{lft:eq:haar}) generalizes to say that any color-nonsinglet average over the gauge group 
vanishes.
Color singlets can propagate, while would-be states with color can be considered to have infinite mass.
Thus, lattice gauge theory confines.

Owing to Eq.~(\ref{lft:eq:haar}), the area law holds for both abelian as well as nonabelian gauge theories.
The salient question, however, is whether the confining behavior persists into the relevant regime of weak
coupling.
In QED, the long-distance coupling in nature is weak, $\alpha=e^2/4\pi=1/137$, and in QCD weak (bare)
coupling corresponds to $a\Lambda_\mathrm{QCD}\ll1$.
In fact, the strong-coupling dynamics of (compact) U(1) lattice gauge theory are influenced by a tangle of
magnetic monopoles~\cite{Savit:1977fw,Banks:1977cc}, unlike what one has in~QED.
The monopoles provide the crucial insight to prove rigorously~\cite{Guth:1979gz,Frohlich:1982gf} that a
first-order phase transition separates the confining, strong-coupling region from a phase with a massless
photon and Coulomb interactions.
The latter phase is QED, while the confining phase of U(1) lattice gauge theory has nothing
to do with~QED.

The rigorous proofs fail in the nonabelian case, however.
At present, there is no accepted rigorous analytic proof that confinement persists for nonabelian theories 
into the weak-coupling regime.
Several analytical and nonperturbative calculations, taken together, provide strong evidence that the
confinement of the strong-coupling limit of lattice gauge theory survives to continuum~QCD.
First, consider how the exponent in Eq.~(\ref{lft:eq:area}) depends on $g_0^2$.
At strongest coupling, Eq.~(\ref{lft:eq:Umn}) implies
\begin{equation}
    \sigma a^2 = \ln[g_0^2(a)/2N],
\end{equation}
whereas at weakest coupling, Eq.~(\ref{lft:eq:LambdaF}) requires
\begin{equation}
    \sigma a^2 \propto e^{-1/\beta_0g_0^2(a)}.
\end{equation}
The issue at hand is whether these two asymptotic behaviors are connected by a smooth function.
Michael Creutz's pioneering numerical calculations~\cite{Creutz:1980zw,Creutz:1980wj} of 
$\langle\tr U_{m\times n}\rangle$ and, hence, $\sigma a^2$, demonstrated a smooth connection between the two
functional forms, with a knee around $g_0^2\approx1$.
Moreover, Pad\'e extrapolations of high-order strong-coupling expansions anticipate the
knee~\cite{Kogut:1979vg,Munster:1980iv}.
These results thus show no evidence for a first-order transition, so the simplest interpretation is that
confinement persists to weak coupling.

The absence of evidence for a phase transition is not the same as evidence for the absence of a phase
transition.
Indeed, numerical studies \emph{do} find first-order transitions in SU($N$) lattice gauge
theory~\cite{Bhanot:1981eb,Bhanot:1981pj}.
To do so, one searches in a general space of lattice couplings, including irrelevant couplings.
In SU($N$) in four dimensions, a line of phase transitions ends, and various trajectories in the space of
couplings smoothly connect the strongly and weakly coupled regimes.
In U(1), the phase-transition line never ends, so the first-order phase transition cannot be circumvented.
It seems unlikely that numerical work has missed a key piece of information about the bulk phase structure of
lattice gauge theory.
The tool's suitability and the community's expertise seem up to the task.

Adding quarks to lattice gauge theory changes the picture of confinement somewhat.
(Lattice-fermion constructions are discussed below.) 
If the source-sink separation is long enough, it is energetically preferable for a quark-antiquark pair to
pop out of the glue and screen the color sink and source.
This behavior can be seen in a double ``high-temperature'' series in $1/g_0^2$ and $1/m_0$.
Terms varying with the area (from the $1/g_0^2$ series) and with the perimeter (from the $1/m_0$ series)
arise, with the former remaining important for small and intermediate separations, and the latter dominating
for large separations.

In addition to the string tension, strong-coupling expansions can be used to compute hadron masses.
QCD is expected to have bound states that lie outside the quark model, such as glueballs, which are composed 
of gluons but no valence quarks.
At leading order in strong coupling, various glueballs are degenerate with common mass
\begin{equation}
    Ma = 4\ln[g_0^2(a)/2N].
\end{equation}
The series have been extended through order $g_0^{-16}$ for scalar ($J^{PC}=0^{++}$), tensor ($2^{++}$), and
axial-vector ($1^{+-}$) glueballs~\cite{Munster:1981es,Seo:1982jh}, yielding ratios~\cite{Smit:1982fx,%
Munster:1982kg} $M_{2^{++}}/M_{0^{++}}\approx1$, $M_{1^{+-}}/M_{0^{++}}=1.8\pm0.3$ (Euclidean spacetime
lattice), with similar results from a continuous time Hamiltonian formulation~\cite{Kogut:1976zr}.

One can also compute meson and baryon masses.
The simplest approach~\cite{Banks:1976ia} takes both $1/g_0^2$ and $1/m_0$ to be small, but the
latter would be far from the up and down quarks.
Another approach is to exploit mean-field theory techniques from statistical mechanics, which permit the
resummation of the $1/m_0$ expansion.
This trick amounts to an expansion in $1/d$, where $d$ is the dimension of spacetime; for $d=4$ the 
expansion parameter is reasonable small.
At strongest coupling, the disorder of the gauge field drives chiral symmetry
breaking~\cite{Blairon:1980pk,Kogut:1982ds}, and the Goldstone boson (pion) mass satisfies 
$M_\pi^2\propto m_0\langle\bar{\psi}\psi\rangle$.
These calculations also find that non-Goldstone meson masses satisfy 
$M\propto\mathrm{const}+\mathrm{O}(m_0)$ and baryon masses (for $N$ colors) satisfy 
$M\propto N\times\mathrm{const}+\mathrm{O}(m_0)$~\cite{KlubergStern:1982bs,Martin:1983hw}.
We shall return to the implications of spontaneous chiral symmetry breaking at the end of 
Sect.~\ref{lft:sec:qcd}.

To end this section, let us discuss the uneasy relationship between fermions and the lattice.\footnote{%
Whence the remark that Heisenberg and Pauli were prudent not to take up the issue.}
(The level of these paragraphs is somewhat higher, and readers can skip them and proceed to 
Sect.~\ref{lft:sec:qcd} without much loss.) 
In the 1974 paper~\cite{Wilson:1974sk}, Wilson used a lattice fermion Lagrangian with (inverse) free
propagator
\begin{equation}
    S^{-1}(p) = ia^{-1}\sum_{\mu=1}^4\gamma_\mu\sin(p_\mu a)+m_0,
    \label{lft:eq:naive}
\end{equation} 
where each component of the momentum $p$ lies in the interval $(-\pi/a,\pi/a]$. 
This expression looks like its continuum counterpart not only for $p\sim0$ but also at the 15 other corners
of the Brioullin zone, $p_\mu\sim 0 \mod \pi/a$.
In the continuum limit, all~16 modes correspond to physical states, which is known as the ``fermion doubling
problem.'' 
The extra states appear everywhere~\cite{Karsten:1980wd}.
For example, they multiply by 16 the fermion-loop contribution to $\beta_0$ [the term proportional to $n_f$
in Eq.~(\ref{lft:eq:beta0})], and they contribute to the axial anomaly with a pattern of signs $1-4+6-4+1=0$ 
(in four dimensions).
The Lagrangian corresponding to Eq.~(\ref{lft:eq:naive}) has an exact chiral symmetry in the massless limit;
hence, the anomaly must vanish in this case (even though this is not desired for QCD).

Several formulations have been introduced to amelioriate the doubling problem.
In a Hamiltonian formulation with discrete space and continuous time (and, hence, only 8 states to start
with), John Kogut and Leonard Susskind~\cite{Kogut:1974ag} put the upper two and lower two components of a
Dirac spinor on the even and odd sites of the lattice, respectively, reducing the number of degrees of
freedom by two.
Susskind~\cite{Susskind:1976jm} later devised a method with one component per site.
A~Euclidean spacetime lattice version of this method~\cite{Kawamoto:1981hw,Sharatchandra:1981si} is now
referred to as staggered fermions.
This formulation exactly preserves a subset of chiral symmetry but still has four fermion states for every
fermion field.
A~non-Noether flavor-singlet axial current is anomalous~\cite{Sharatchandra:1981si,Smit:1987zh}.

Wilson~\cite{Wilson:1977nj} introduced a dimension-five term that yields a large mass to the 15 extra states.
The axial anomaly is obtained correctly, which is possible because the Wilson term breaks the axial
symmetries.
In practice, one has a fine-tuning problem here: the mass term and the Wilson term must balance each other to
provide the small amount of explicit axial-symmetry breaking of QCD.
After this fine-tuning, which can be carried out nonperturbatively, the residual chiral-symmetry breaking is
proportional to the lattice spacing.
One can add to the action a Pauli term~\cite{Sheikholeslami:1985ij}, which is also of dimension five, and
then impose Ward identities~\cite{Jansen:1995ck,Luscher:1996sc} to reduce discretization effects to O($a^2$).
For two flavors of Wilson fermions, it is also possible to remove the leading-order discretization effect via
an isospin off-diagonal mass term~\cite{Frezzotti:2000nk,Frezzotti:2003ni}, which is known as ``twisted
mass.''

On a lattice, chiral symmetry and the doubling problem are deeply connected, which is encapsulated in the
Nielsen-Ninomiya theorem~\cite{Nielsen:1980rz,Friedan:1982nk}.
A~way around this theorem comes from the Ginsparg-Wilson relation~\cite{Ginsparg:1981bj}, which uses 
renormalization-group ideas to derive a minimal condition on lattice chiral symmetry (for Dirac fermions).
The Ginsparg-Wilson relation reads
\begin{equation}
    \gamma_5D + D\gamma_5=aD\gamma_5D,
\end{equation}
where $D$ is the lattice Dirac operator.
The Nielsen-Ninomiya theorem assumes the right-hand side vanishes and, thus, does not apply.
Solutions to the Ginsparg-Wilson relation~\cite{Shamir:1993zy,Neuberger:1997fp,Hasenfratz:1998jp} are
compatible with a suitably modified chiral transformation~\cite{Luscher:1998pqa} but are computationally more
demanding than the other methods.
This setup allows a rigorous derivation of soft-pion theorems~\cite{Chandrasekharan:1998wg}.
These ideas are also closely~\cite{Kaplan:1992bt,Narayanan:1994gw} or intimately~\cite{Luscher:1998du,%
Luscher:1999un} related to ideas to formulate chiral gauge theories (such as
the standard electroweak interaction) on the lattice.

Staggered~\cite{Susskind:1976jm,Kawamoto:1981hw,Sharatchandra:1981si}, twisted-mass
Wilson~\cite{Frezzotti:2000nk,Frezzotti:2003ni}, improved Wilson~\cite{Sheikholeslami:1985ij},
domain-wall~\cite{Shamir:1993zy}, and overlap~\cite{Neuberger:1997fp} fermions are all used in the
large-scale computations discussed below.

\section{The Origin of (Your) Mass}
\label{lft:sec:qcd}

Although strong-coupling expansions provided new insight into gauge theories, it became clear that they 
would not offer a path to small, robust error bars.
Today, a set of numerical Monte Carlo techniques are the largest focus of research in lattice gauge theory.  
In many cases, for example, the computation of hadronic matrix elements in electroweak processes, the goal 
is to provides a solid number with a full error budget.
To understand mass, however, one would like to have more than numbers, but also a qualitative understanding.
As we shall see, numerical calculations have played a key role here too.

Let us begin with a short explanation of the numerical methods.
In all cases of interest, the action in Eq.~(\ref{lft:eq:vev}) can be written
$S=S_\mathrm{gauge}+\bar{\psi}(D+m)\psi$, where $D$ is a matrix with spacetime, color, flavor, and Dirac
indices, and $m$ is a mass matrix (diagonal in all indices).
To obtain a nonzero result, the number of fermion and antifermion fields in Eq.~(\ref{lft:eq:vev}) must be
the same.
Suppose the number is $A$; that means that the product of $\Phi$s can be re-expressed as
\begin{equation}
    \prod_{i=1}^n \Phi_i = \phi(U)\prod_{a=1}^A \bar{f}_a(U)\psi\,\bar{\psi}f_a(U),
\end{equation}
where $\bar{f}_a(U)$ and $f_a(U)$ account for all structure attached to fermions and antifermions on the
left-hand side, and $\phi(U)$ stands for whatever remains.
To calculate hadron masses, we need two-point functions ($n=2$ on the left-hand side) for mesons ($A=2$ on
the right-hand side) and baryons ($A=3$).
These two-point functions can be expressed as ($x_4>0$)
\begin{equation}
    \left\langle\Phi^\dagger_i(x_4)\Phi_j(0)\right\rangle = \sum_{r=0}^\infty 
        \left\langle0\left|\hat{\Phi}^\dagger_i\right|r\right\rangle
        \left\langle r\left|\hat{\Phi}_j\right|0\right\rangle
        \exp\left(-M_rx_4\right),
    \label{lft:eq:transfer}
\end{equation}
where the $\Phi_i$ have specific three-momentum and flavor quantum numbers, 
and~$M_r$ is the energy of the $r$th radial excitation with the quantum numbers of~$\Phi_i$.
For three-momentum $\bm{p}=\bm{0}$, energy means mass.
For simple actions, Eq.~(\ref{lft:eq:transfer}) is a theorem~\cite{Wilson:1973jj,Creutz:1976ch,%
Luscher:1976ms,Osterwalder:1977pc,Sharatchandra:1981si} and, for more general lattice actions, essentially a 
theorem~\cite{Luscher:1984is}.
Given the left-hand side from a numerical computation, the masses are extracted by fitting the 
numerical data to the right-hand side.
These fits can be improved by choosing $x_4$ large enough to suppress highly excited states and by choosing 
the~$\Phi_i$ to project mostly onto one specific state.

The integration over fermionic variables can be carried out by hand, yielding
\begin{eqnarray} \hspace*{-2em}
    \left\langle\phi(U)\prod_{a=1}^A \bar{f}_a\psi\,\bar{\psi}f_a \right\rangle & = &
        \frac{1}{Z} \int \prod_{x,\mu} dU_{x,\mu} \; \phi(U)\, 
        \det_{a,b}\left\{\bar{f}_a[D(U)+m]^{-1}f_b\right\}
        \times \nonumber \\ & & \hspace*{11em} 
        \Det[D(U)+m] \, e^{-S_\mathrm{gauge}(U)}, 
    \label{lft:eq:vevDet}    
\end{eqnarray}
where $\det_{a,b}$ is a normal determinant over the enumeration of fermion fields, 
while $\Det$ denotes a determinant over spacetime, color, flavor, and Dirac indices.
Physically, $\det_{a,b}\bar{f}_a [D(U)+m]^{-1}f_b$ represents the propagators of valence quarks in the 
$n$-point function, while $\Det[D(U)+m]$ denotes sea quarks---virtual quark-antiquark pairs bubbling out of 
the stew of gluons.

The number of independent variables of integration is huge if the spatial extent is to be larger than a
hadron and the lattice spacing much smaller than a hadron.
The only feasible numerical technique for computing such integrals is a Monte Carlo method with importance
sampling.
That means to generate $C$ configurations of $\{U_{x,\mu},~\forall\,x,\,\mu\}^{(c)}$ chosen randomly with
weight $\Det[D(U)+m] \, e^{-S_\mathrm{gauge}(U)}$.
Then
\begin{equation}
    \left\langle\phi(U)\prod_a \bar{f}_a\psi\,\bar{\psi}f_a \right\rangle = \lim_{C\to\infty}
        \frac{1}{C} \sum_{c=1}^{C} \phi(U^{(c)})\, 
        \det_{a,b}\left\{\bar{f}_a[D(U^{(c)})+m]^{-1}f_b\right\}.
    \label{lft:eq:important}    
\end{equation}
In practice, $C$ is finite but as large as possible.
The details of the numerical algorithms lies beyond the scope of this article;
for a pedagogical review, see Ref.~\onlinecite{DiPierro:2005vz}.

The second-most computationally demanding part of this procedure is to obtain the valence-quark propagators
$\bar{f}_a[D(U)+m]^{-1}f_b$.
The most demanding part is to account for the sea-quark factor $\Det[D(U)+m]$ in the importance sampling.
Early mass calculations thus used a valence approximation~\cite{Weingarten:1981jy}, computing each
$\bar{f}_a[D(U^{(c)})+m]^{-1}f_b$ while replacing $\Det[D(U^{(c)})+m]$ with~1.
In addition, the elimination of bare parameters in favor of physical quantities absorbs an implicitly
specified part of the physical effects of the sea quarks.
The valence approximation is better known as the quenched approximation, from an analogy with
condensed-matter systems~\cite{Marinari:1981qf}.

\begin{figure}[bp]
    \centering
    \includegraphics[width=0.5\textwidth]{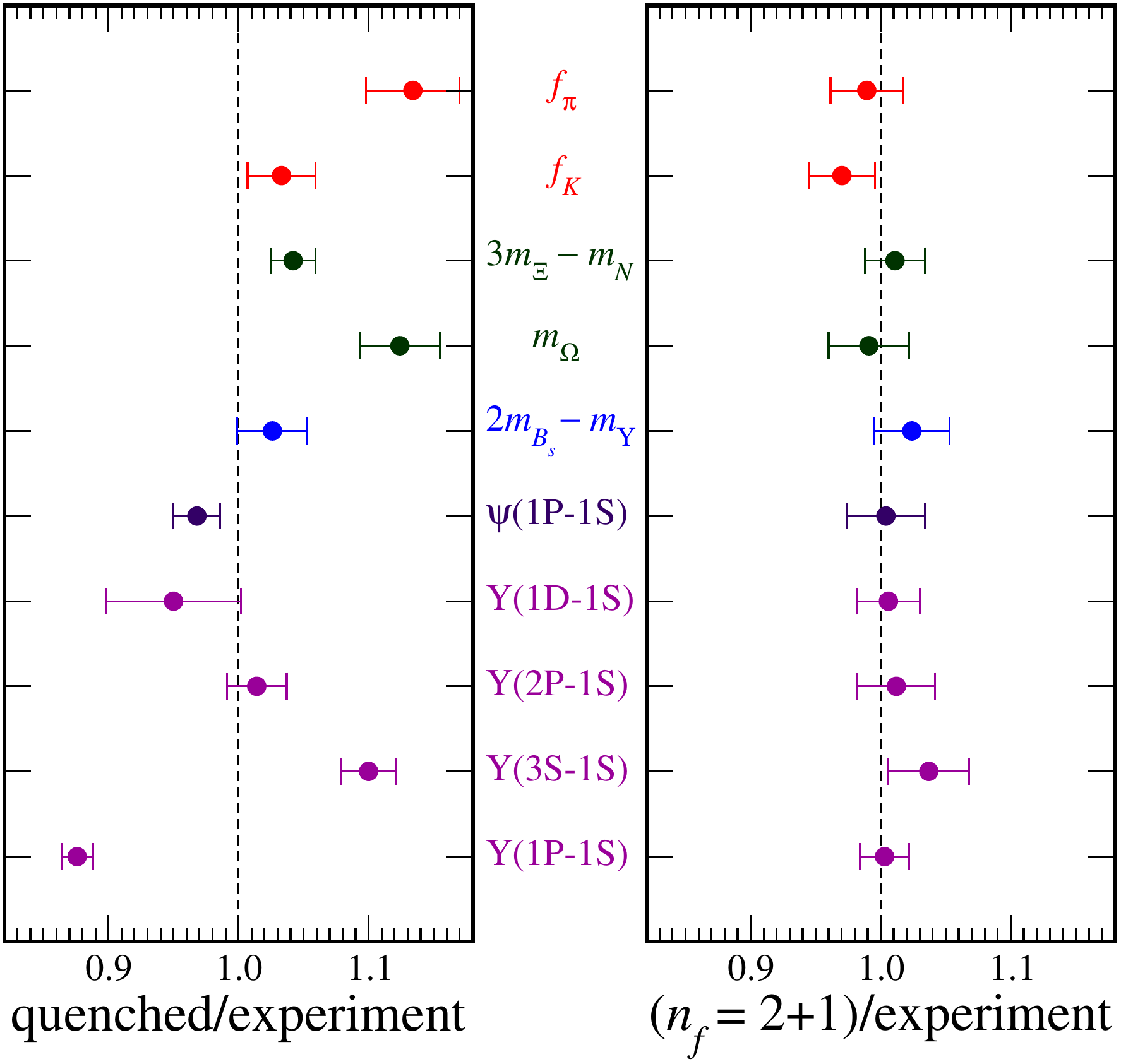}
    \caption[lft:fig:ratio]{Comparison of quenched and $2+1$ lattice-QCD calculations, showing only 
        statistical errors (to assess the systematics of quenching)~\cite{Davies:2003ik}.
        Free parameters $(g_0^2(a), \half(m_u+m_d), m_s)$ fixed with the 2S-1S splitting of bottomonium,
        $M_\pi$, and~$M_K$.}
    \label{lft:fig:ratio}
\end{figure}

There are too many quenched calculations of hadron masses in the literature to provide a useful survey.
Nowadays, the inclusion of sea quarks is commonplace.
Indeed, in some important applications, such as the thermodynamics of QCD, the sea quarks play an absolutely
crucial role.
The remainder of this article focuses, therefore, on numerical calculations that include the sea of up, down,
and strange quarks; these are usually called 2+1-flavor calculations.
In some cases, charmed sea quarks are included as well, and these are known as 2+1+1-flavor calculations.

Figure~\ref{lft:fig:ratio} shows a comparison of quenched and 2+1 calculations for a wide variety of 
masses and pseudoscalar-meson decay constants~\cite{Davies:2003ik}.
The results of the lattice-QCD calculations are divided by their corresponding entries in the (2002 edition
of the) review of particle physics from the Particle Data Group (PDG)~\cite{Hagiwara:2002fs}.
Results should ideally lie close to~1.
The quenched results lie with 10--15\% of PDG values, sometimes closer, but the pattern of (nonstatistical)
variation is hard to understand.
Upon adding 2+1 flavors of sea quarks, the discrepancies disappear.

Computational science often develops in a way that festoons the basics with many specialized methodological
improvements.
Nonexperts often react by putting the whole process into a black box to shield themselves from the details.
They are then comforted by genuine predictions: calculations for which the correct result was not known in
advance, but which are then confirmed by other means, e.g., experimental measurements.

Soon after the publication of in Fig.~\ref{lft:fig:ratio}, lattice QCD enjoyed several such predictions,
including the shape of form factors in semileptonic $D$ decays~\cite{Aubin:2004ej}, the mass of the $B_c$
meson (composed of a bottom quark and a charmed antiquark)~\cite{Allison:2004be}, the decay constants of
charmed mesons~\cite{Aubin:2005ar}, and the mass of the $\eta_b$ meson (the lightest bottomonium
state)~\cite{Gray:2005ur}.
\begin{figure}[bp]
    \centering
    \includegraphics[width=0.48\textwidth]{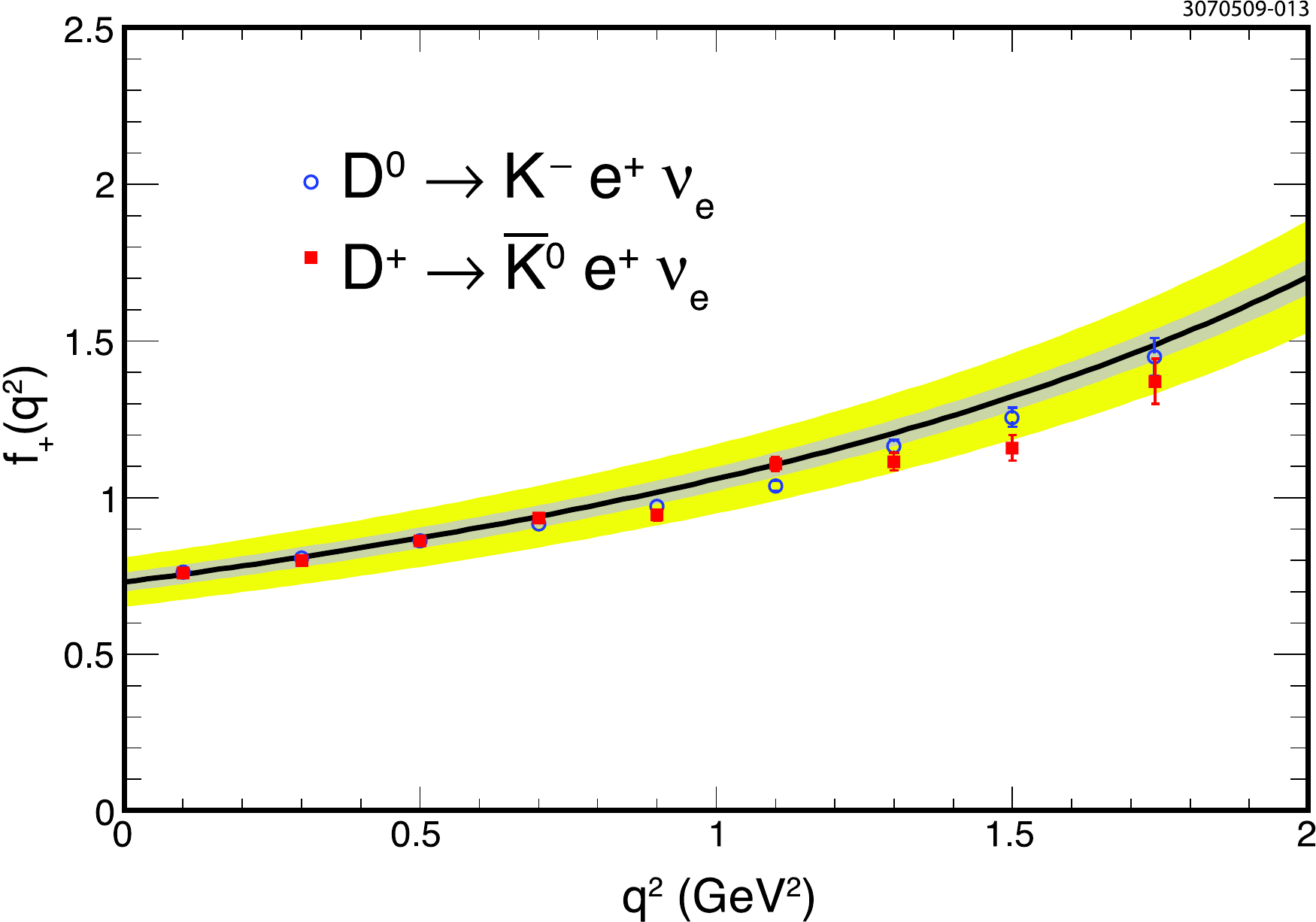}\hfill
    \includegraphics[width=0.48\textwidth]{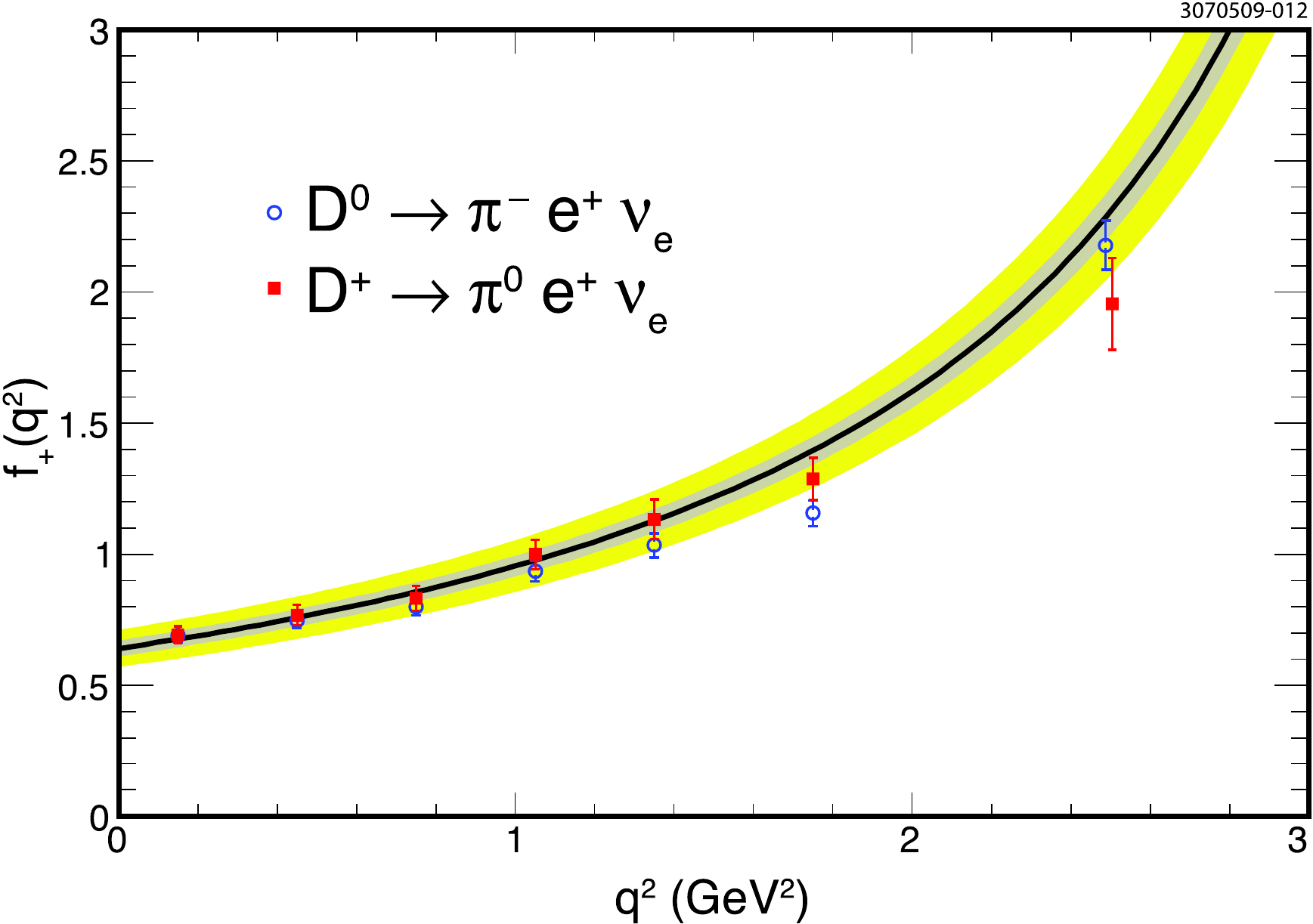}
    \caption[lft:fig:ratio]{Comparison~\cite{Besson:2009uv} of $2+1$ lattice-QCD calculations of $D$-meson 
        form factors~\cite{Aubin:2004ej,Bernard:2009ke} (curves with error bands) with 
        measurements from CLEO~\cite{Besson:2009uv} (points with error bars).}
    \label{lft:fig:D2K}
\end{figure}
Figure~\ref{lft:fig:D2K} shows measurements of the form factors for $D\to K\ell\nu$ and $D\to\pi\ell\nu$ by
the CLEO experiment~\cite{Besson:2009uv} overlaying a lattice-QCD calculation~\cite{Aubin:2004ej} with an
improved visualization of systematic errors~\cite{Bernard:2009ke}.
The CLEO data~\cite{Besson:2009uv} are the most precise among several experiments confirming the lattice-QCD
calculation; cf.~FOCUS~\cite{Link:2004dh}, Belle~\cite{Widhalm:2006wz}, BaBar~\cite{Aubert:2007wg}, 
and earlier CLEO measurements~\cite{Dobbs:2007aa,Ge:2008aa}.

\begin{figure}[bp]
    \includegraphics[width=\textwidth]{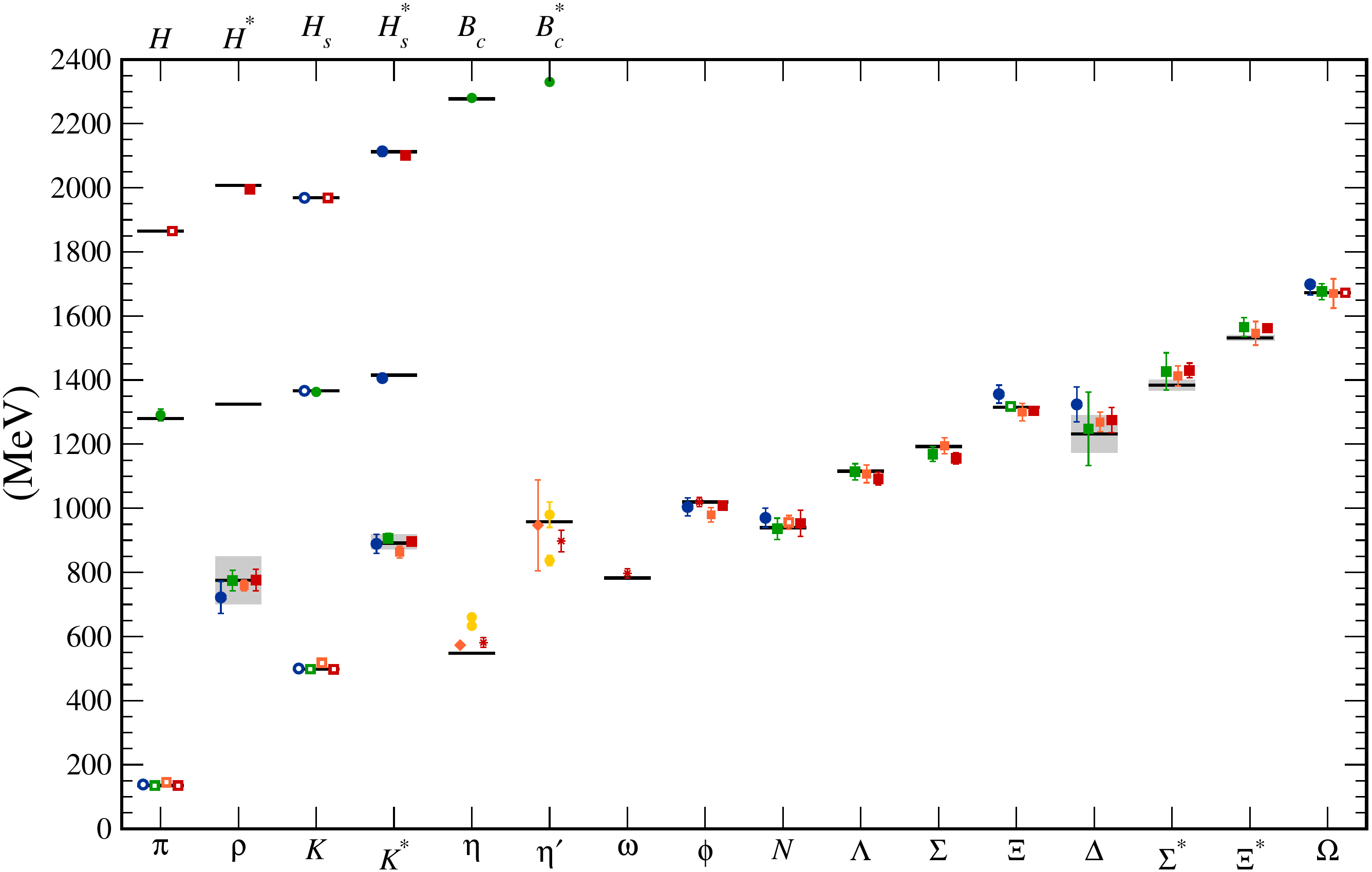}
    \caption[lft:fig:spectrum]{Hadron mass spectrum with 2+1 flavors of sea quarks,
        from Ref.~\onlinecite{Kronfeld:2012uk}.
        Results for many light mesons and baryons are from 
        MILC~\cite{Aubin:2004wf,Bazavov:2009bb},
        PACS-CS~\cite{Aoki:2008sm},
        BMW~\cite{Durr:2008zz}, and
        QCDSF~\cite{Bietenholz:2011qq}.
        Results for the $\eta$ and $\eta'$ mesons are from 
        RBC \& UKQCD~\cite{Christ:2010dd},
        Hadron Spectrum~\cite{Dudek:2011tt}, and
        UKQCD~\cite{Gregory:2011sg}.
        Result for the $\omega$ meson is from Hadron Spectrum~\cite{Dudek:2011tt}.
        Results for heavy-light mesons are from 
        Fermilab-MILC~\cite{Bernard:2010fr},
        HPQCD~\cite{Gregory:2010gm}, and
        Mohler \& Woloshyn~\cite{Mohler:2011ke}.
        $b$-flavored meson masses are offset by $-4000$~MeV.
        Circles, squares, and diamonds denote staggered, Wilson, and chiral sea quarks, respectively. 
        Asterisks represent anisotropic lattices, $a_4/a_i<1$.
        Open symbols denote inputs; filled symbols and asterisks denote output results.
        Red, orange, yellow, green, and blue denote increasing numbers of ensembles 
        (i.e., range of lattice spacing and depth of sea quark masse).
        Horizontal bars (gray boxes) denote experimentally measured masses (widths).}
    \label{lft:fig:spectrum}
\end{figure}

Before turning to hadron-mass calculations, let us take stock of the numerical results discussed so far.
Section~\ref{lft:sec:lattice} noted that Monte Carlo calculations of simple quantities such as Wilson loops
agreed with both strong-coupling and weak-coupling expansions, in their respective domains of applicability.
Here, we have seen that quarkonium masses and some other properties of heavy-quark systems---as well as
leptonic decay constans $f_\pi$ and~$f_K$---agree very well with experimental measurements, even when those
were not known ahead of time.
When combined with the numerous self-consistency checks that every modern large-scale lattice-QCD calculation
undergoes, it is fair to say that the techniques for generating and analyzing numerical data have matured.
In particular, the standards for estimating full error budgets have become, by and large, high.

With confidence bolstered by these remarks, let us now examine recent calculations of the hadron mass 
spectrum.
A summary is shown in Fig.~\ref{lft:fig:spectrum}.
More details about the underlying work can be found in the review from which this plot is
taken~\cite{Kronfeld:2012uk} or in a comprehensive review of hadron mass
calculations~\cite{Fodor:2012gf}.
The most important features are as follows.
Many different groups of researchers (symbol shape and color) have carried out these calculations,
and they all find broad agreement with nature.
They use different fermion formulations (symbol shape) and a different range of lattice spacing and quark
masses (symbol color).
The total errors in many cases are small.
In particular, the nucleon mass---the main contributor to everyday mass---has an error of around~$2\%$.

\begin{figure}[bp]
    \includegraphics[width=\textwidth]{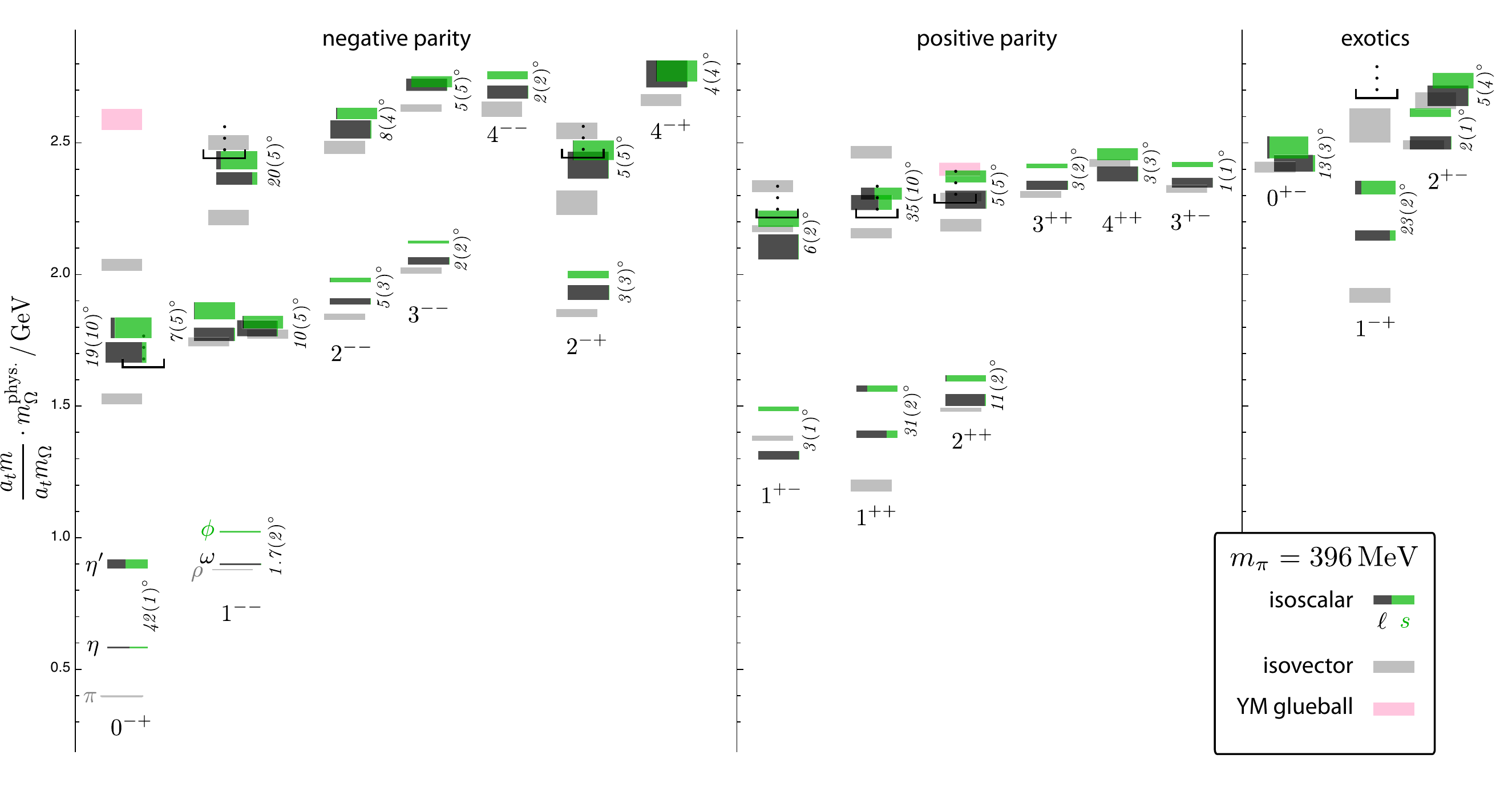}
    \caption[lft:fig:excited]{Excited-state meson spectrum~\cite{Dudek:2011tt}, including 
        isoscalar-$\bar{s}s$ mixing (shades of green) and pure-gauge glueballs (pink).}
    \label{lft:fig:excited}
\end{figure}

Figure~\ref{lft:fig:spectrum} shows only the lowest-lying state in each channel, cf.\ 
Eq.~(\ref{lft:eq:transfer}).
Excited states pose more technical challenges, starting with a lower signal-to-noise ratio in the Monte 
Carlo estimates of the two-point functions.
Nevertheless, recent progress in this area has been encouraging.
An example for mesons is shown in Fig.~\ref{lft:fig:excited}.
A~further noteworthy feature of these calculations is that the same techniques that disentangle the
excited-state spectrum also yield mixing angles.
These mixing angles agree well with corresponding experimental measurements~\cite{Beringer:1900zz}.
Figure~\ref{lft:fig:excited} also shows results for exotic mesons, which have $J^{PC}$ quantum numbers that
are inaccessible from the quark model, and for glueballs (in the quenched approximation~%
\cite{Morningstar:1999rf}, though more recent glueball calculations with 2+1 sea quarks find similar
glueball masses~\cite{Richards:2010ck}).
These glueball masses validate the axial-vector/scalar but not the tensor/scalar ratios of strong coupling.
Excited baryon mass calculations have also been carried out~\cite{Bulava:2009jb,Edwards:2011jj} and will be
tested by experiments at Jefferson Laboratory~\cite{Burkert:2009zf}.

Figure~\ref{lft:fig:spectrum} shows that we have obtained a solid, quantitative understanding of the mass of
simple hadrons, and Fig.~\ref{lft:fig:excited} shows that this understanding is improving for more
complicated hadrons.
To understand how the mass is generated, however, a qualitative understanding is also necessary.
Here, let us return to the valence approximation to obtain a physical picture.

In electrodynamics, one measures the force by varying the distance between two static charges.
The force arises from the energy stored in the electric dipole field between the two charges.
As a quantum system, the field actually has discrete energy levels, the lowest being the (semiclassical)
potential energy (whose gradient yields the force).
The same holds for the chromoelectric dipole field between a static source and static sink of color.
Now, however, the shape of the dipole field is influenced by the gluon self-interaction: chromoelectric field
lines attract each~other.

\begin{figure}[bp]
    \includegraphics[width=\textwidth]{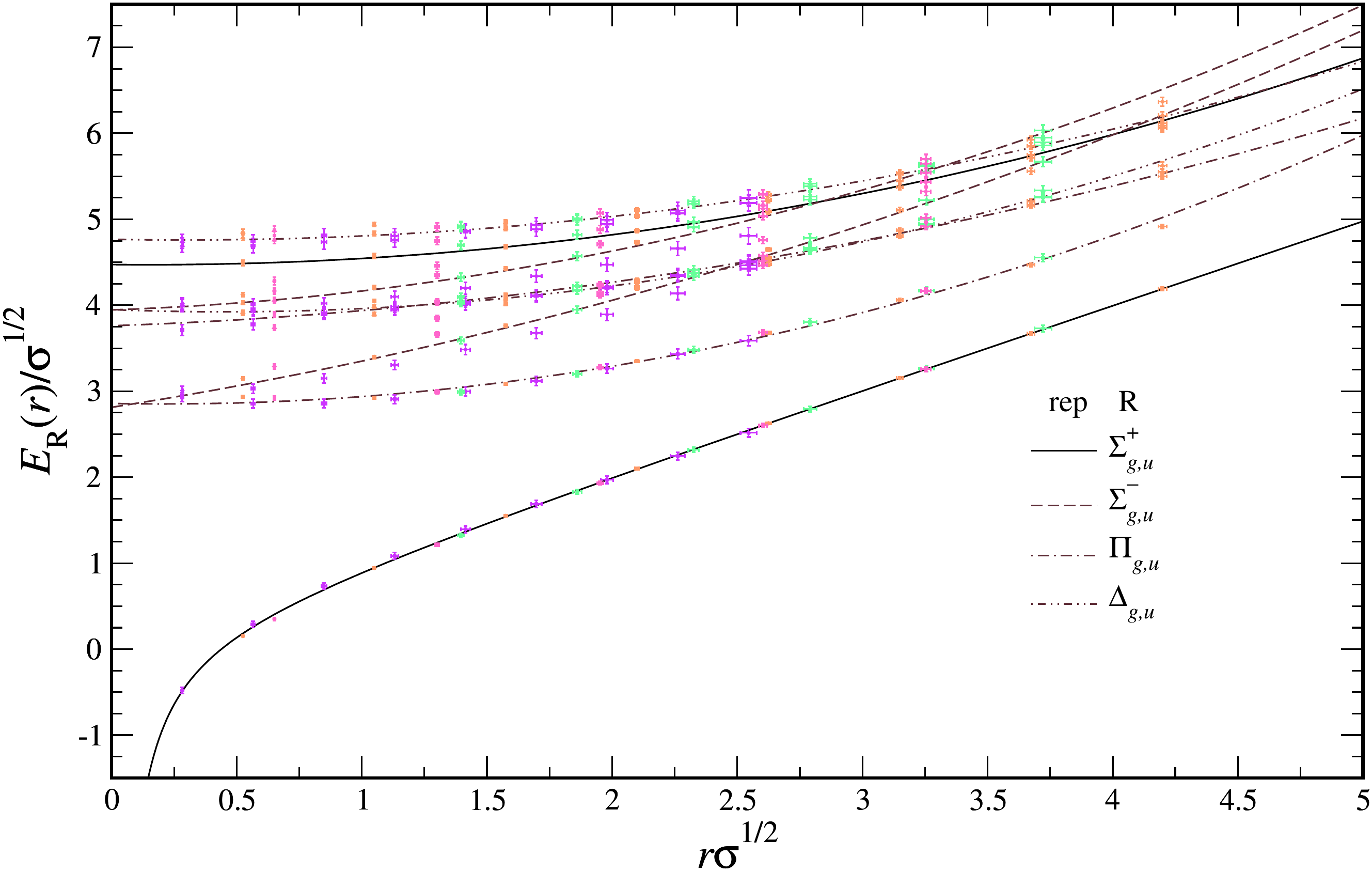}
    \caption[lft:fig:excited]{Excited-state spectrum of potentials~\cite{Juge:2002br}.
        The representations $R$ correspond to the gluonic angular momentum along the source-sink axis, with 
        subscript $g$ ($u$) for $CP=\pm1$, and for $\Sigma$ wave a superscript for parity upon reflection 
        through the midpoint.
        To convert from string-tension units to physical units, note 
        $\sigma^{1/2}\approx420~\text{MeV}\approx2.1~\textrm{fm}^{-1}$.
        Plot from Ref.~\onlinecite{Kronfeld:2012uk}.}
    \label{lft:fig:sausage}
\end{figure}

Figure~\ref{lft:fig:sausage} shows the excitation spectrum of the chromoelectric dipole
field~\cite{Juge:2002br}.
At short distances, the level spacing and ordering is in accord with asymptotic freedom.
For example, the lowest level, $V(r)$ is Coulombic up to logarithmic corrections.
As~$r$ increases, the spacing between the levels changes.
At a separation of around 2~fm, the level ordering rearranges to that of a string, but the level spacing does
not become fully string-like until larger separations~\cite{Juge:2004xr}.
At large distances, the potential $V(r)$ becomes linear in $r$; this behavior is the area law of
Eq.~(\ref{lft:eq:area}), now, however, near the continuum limit and only at large enough distances.
A~vivid picture of the flux tube has it narrowing as $r$ increases, owing to the attraction between gluons,
but the details suggest that the flux tube retains diameter a bit less than 1~fm \cite{Luscher:1980iy}.
At hadronic length scales, it looks more like a sausage than a string.
This picture holds even when quarks are added to the calculations: the linear region persists out to around 
1.25~fm ($r\sigma^{1/2}\approx2.5$), at which point the flux tube breaks~\cite{Bali:2005fu}.

The linear rise of the potential for $r\gtrsim0.4$~fm provides a striking explanation of the hadron
masses~\cite{Greensite:2003bk}.
The energy stored in a flux-tube of length $L$ and string tension $\sigma$ is simply~$\sigma L$.
(Here, $L$ should be large enough to be in the linear regime but shorter that the string-breaking distance.)
Imagine attaching a massless quark to one end of the tube and an antiquark to the other.
The ends must move with the speed of light, and the total mass $M=E/c^2$ of this ``meson'' is
\begin{equation}
    M = \int_{-L/2}^{L/2} \frac{\sigma\,dx}{[1-v(x)^2]^{1/2}}   = 
        \int_{-L/2}^{L/2} \frac{\sigma\,dx}{[1-(2x/L)^2]^{1/2}} = \half\pi\sigma L,
    \label{lft:eq:ReggeM}
\end{equation}
where the denominator accounts for relativistic motion.
The kinetic energy increases the mass by $\frac{\pi}{2}-1\approx60\%$.
The angular momentum of this system is
\begin{equation}
    J = \int_{-L/2}^{L/2} \frac{\sigma\,v(x)x\,dx}{[1-v(x)^2]^{1/2}}   = \frac{2}{L}
        \int_{-L/2}^{L/2} \frac{\sigma\,x^2\,dx}{[1-(2x/L)^2]^{1/2}} = 
    \eighth\pi\sigma L^2 = \frac{M^2}{2\pi\sigma},
    \label{lft:eq:ReggeJ}
\end{equation}
where the last step comes from eliminating $L$ in favor of~$M$.
Experimental measurements of meson masses and spin satisfy such linear relationships---known as Regge
trajectories---between $J$ and~$M^2$, albeit with nonzero intercepts (from effects neglected here).
Equations~(\ref{lft:eq:ReggeM}) and~(\ref{lft:eq:ReggeJ}) are interesting because they are simple.
The idea behind them is supported, however, by the empirical observation that heavy-light meson wave
functions computed directly with lattice gauge theory coincide with Schr\"odinger wave functions determined
from a relativistic kinetic energy and the heavy-quark potential computed with lattice gauge
theory~\cite{Duncan:1992eb}.

Let us return to chiral symmetry breaking.
Before quarks or partons had been proposed, Nambu~\cite{Nambu:1960xd} pointed out that the small mass of the
pion (140~MeV) could be explained if an axial symmetry was spontaneously broken.
QCD possesses such symmetries in the limit of vanishing quark mass.
Indeed, in this idealization, the pion mass would vanish by the following theorem~\cite{Goldstone:1961eq}
\begin{equation}
    M_\pi^2 \langle\bar{\psi}\psi\rangle = 0.
\end{equation}
This picture has been demonstrated via quantiative lattice-QCD calculations of the chiral
condensate~\cite{Fukaya:2010na}, firmly establishing $\langle\bar{\psi}\psi\rangle\neq0$.
The nonzero pion mass arises owing to the explicit symmetry breaking from the up and down quark masses.
As a consequence, one expects $M_\pi^2\propto m_q$, which has been amply demonstrated in lattice
QCD~\cite{Bazavov:2009bb,Fodor:2012gf}.
Since the nucleon and pion experience residual strong interactions, the nucleon is surrounded with a cloud of
pions.
The size of the nucleon, and other hadrons is, thus, dictated by the pion Compton wavelength; the density of
nuclear material is proportional to~$M_NM_\pi^3$.

The richness of everyday life stems from chemistry, which, in turn, hinges both on an attractive force
between protons and neutrons to hold atomic nuclei together, and a short-range repulsive force to aid nuclear
stability.
In QCD, the attractive force is akin to van~der~Waals forces among molecules and can be vividly and
successfully modeled by meson exchange, particularly pion exchange.
The detailed, first-principles study of these forces is just beginning~\cite{Ishii:2006ec,Beane:2012vq}.
Recent developments have been encouraging and illustrate that the origin of mass is not the only exciting 
problem in physics.

\section{Summary and Outlook}
\label{lft:sec:end}

The origin of mass is a compelling problem with many facets.
This article has touched on only one, the origin of mass of everyday objects, which can be pinpointed
directly to the protons and neutrons in atomic nuclei.
Remarkably, most of the nucleon mass has a dynamical origin: strong confining forces influenced by chiral
symmetry breaking generate the mass and size of nucleons and, hence, nuclei.
We understand these dynamics as quantum chromodynamics.
With powerful numerical calculations based on lattice gauge theory, we have disentangled puzzles and verified
many conjectures.

Because of asymptotic freedom, QCD as a quantum field theory holds consistently at all energy scales.
That said, as the exploration of particles physics unfolds in the future, it is conceivable that physicists
will discover a substructure to quarks or a unification of the chromodynamic interaction with the other gauge
interactions of the standard model.
Such discoveries would relegate the SU(3) gauge symmetry of QCD to a (relatively) low-energy description of
nature.
Moreover, in such frameworks a high-energy value of the QCD gauge coupling is specified, and, in many cases,
a set of thresholds affecting its running is specified as well.
One can thus imagine connecting $\Lambda_\mathrm{QCD}$ to the scales of a more fundamental, more microscopic
theory of (most) everything.
Even so, one would still have to concede that chromodynamics generate everyday mass.
The key physics is the attraction of gluons to each other, the relativistic kinetic energy of light quarks,
and the constraints imposed by dynamical chiral symmetry breaking.

\section*{Acknowledgments}

My sense of the history of lattice gauge theory has been shaped by conversations over the years with 
Bernd Berg,
Robert Finkelstein, 
Chris Hill, 
Peter Lepage, 
Paul Mackenzie, 
Gernot M\"unster, 
Don Petcher, 
Junko Shigemitsu,
Don Weingarten, and 
Ken Wilson, 
among others.
In particular, Gernot M\"unster told me about Werner Heisenberg and Piet Hein.

Fermilab is operated by Fermi Research Alliance, LLC, under Contract No.~DE-AC02-07CH11359 with the United
States Department of Energy.
I thank the Galileo Galilei Institute for Theoretical Physics for hospitality, and the INFN for partial
support, while this article was being completed.

\section*{Appendix: Heisenberg's \emph{Gitterwelt} and Hein's Soma}
\label{lft:sec:kook}

Werner Heisenberg hoped for more from the lattice than mere mathematical rigor.
In a 1930 letter to Niels Bohr, he argued that a universe with a fundamental length, such as a spatial
lattice spacing, would not suffer from many problems (then) facing quantum field theory and nuclear and
atomic physics.
For a translation of the letter and reconstruction of Heisenberg's ideas, see Carazza and
Kragh~\cite{Carazza:1995lt}.
Bohr responded disapprovingly to the idea.
Heisenberg did not publish a paper on his ``\emph{Gitterwelt}'' (``lattice world''), as it came to be known,
although he did make a technical remark that the lattice tames the ultraviolet divergence in the electron's
self energy~\cite{Heisenberg:1930se}.
Nevertheless, Heisenberg's \emph{Gitterwelt} developed a philosophical and scientific following, which was
met with some disdain~\cite{Carazza:1995lt,Kragh:1995am}.
I've been told~\cite{Hill} that when Wilson presented his lattice gauge theory in a seminar at Caltech, he
deflected an aggressive line of questioning from Richard Feynman with,
``I am not a kook; this is not a kook's lattice!'' 
This give-and-take seems to reflect a lingering apprehension against a lattice as fundamental, while 
underappreciating its mathematical utility.

\begin{figure}
    \centering
    \includegraphics[width=0.8\textwidth]{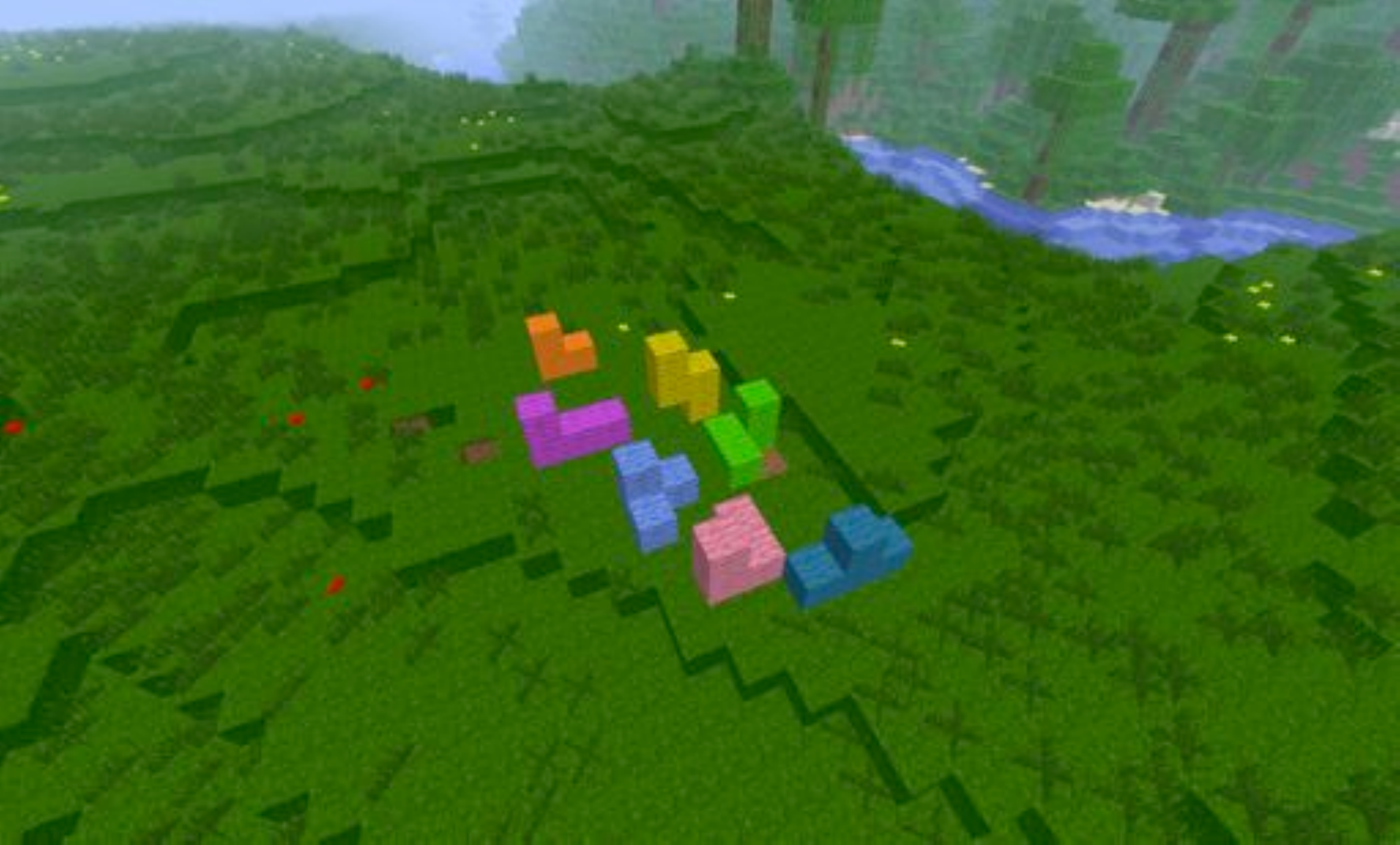}
    \caption[lft:fig:soma]{The seven shapes in Soma grew out of lattice field theory.
        Graphic by Alexander Kronfeld created with Minecraft (\copyright\ 2009--2012 Mojang).}
    \label{lft:fig:soma}
\end{figure}

A lasting outgrowth of Heisenberg's lattice world lies not in theoretical physics but in a geometric puzzle
called Soma, which was created by the Danish inventor and poet Piet Hein~\cite{Soma,Gardner:1961}.
Sometime in the early 1930s, Hein---among other avocations a physics groupie---attended a lecture by
Heisenberg in Copenhagen.
Whether the lecture was on the quantization of QED or on the lattice world, no one seems to know.
Bored, Hein sketched a small three-dimensional lattice on a piece of paper and realized something interesting.
The seven irregular shapes made from three or four cubes (see Fig.~\ref{lft:fig:soma}) can be assembled into
a larger $3\times3\times3$ cube.
These pieces can be assembled in many other mind-bending ways, and Soma has become one of the most popular
three-dimensional puzzles of all time.

\bibliographystyle{ws-rv-van}
\bibliography{history}

\begin{thebibliography}{157}
\providecommand{\natexlab}[1]{#1}
\providecommand{\url}[1]{\texttt{#1}}
\expandafter\ifx\csname urlstyle\endcsname\relax
  \providecommand{\doi}[1]{doi: #1}\else
  \providecommand{\doi}{doi: \begingroup \urlstyle{rm}\Url}\fi

\bibitem{Aad:2012qt}
G.~Aad et~al., Observation of a new particle in the search for the standard
  model {Higgs} boson with the {ATLAS} detector at the {LHC}, \emph{Phys.
  Lett.} {\bf B716}, \penalty0 1--29,  (2012).
\newblock \doi{10.1016/j.physletb.2012.08.020}.

\bibitem{Chatrchyan:2012qt}
S.~Chatrchyan et~al., Observation of a new boson at a mass of {125~GeV} with
  the {CMS} experiment at the {LHC}, \emph{Phys. Lett.} {\bf B716}, \penalty0
  30--61,  (2012).
\newblock \doi{10.1016/j.physletb.2012.08.021}.

\bibitem{Aaltonen:2012qt}
T.~Aaltonen et~al., Evidence for a particle produced in association with weak
  bosons and decaying to a bottom-antibottom quark pair in {Higgs} boson
  searches at the {Tevatron}, \emph{Phys. Rev. Lett.} {\bf 109}, \penalty0
  071804,  (2012).
\newblock \doi{10.1103/PhysRevLett.109.071804}.

\bibitem{Thomson:1904pp}
J.~J. Thomson, On the structure of the atom, \emph{Phil. Mag.} {\bf 7},
  \penalty0 237--265,  (1904).
\newblock \doi{10.1080/14786441308635024}.

\bibitem{Nagaoka:1904sr}
H.~Nagaoka, Kinetics of a system of particles illustrating the line and the
  band spectrum and the phenomena of radioactivity, \emph{Phil. Mag.} {\bf 7},
  \penalty0 445--455,  (1904).
\newblock \doi{10.1080/14786440409463141}.

\bibitem{Geiger:1909ab}
H.~Geiger and E.~Marsden, On a diffuse reflection of the $\alpha$-particles,
  \emph{Proc. Roy. Soc. Lond. A}. {\bf 82}, \penalty0 495--500,  (1909).
\newblock \doi{10.1098/rspa.1909.0054}.

\bibitem{Rutherford:1911zz}
E.~Rutherford, The scattering of $\alpha$ and $\beta$ particles by matter and
  the structure of the atom, \emph{Phil. Mag.} {\bf 21}, \penalty0 669--688,
  (1911).
\newblock \doi{10.1080/14786440508637080}.

\bibitem{Chadwick:1932ma}
J.~Chadwick, Possible existence of a neutron, \emph{Nature}. {\bf 129},
  \penalty0 312,  (1932).
\newblock \doi{10.1038/129312a0}.

\bibitem{Feynman:1969ej}
R.~P. Feynman, Very high-energy collisions of hadrons, \emph{Phys. Rev. Lett.}
  {\bf 23}, \penalty0 1415--1417,  (1969).
\newblock \doi{10.1103/PhysRevLett.23.1415}.

\bibitem{Feynman:1972pm}
R.~P. Feynman, \emph{Photon-Hadron Interactions}. (Benjamin, Reading, MA,
  1972).

\bibitem{Zweig:1981pd}
G.~Zweig.
\newblock An {SU(3)} model for strong interaction symmetry and its breaking.
\newblock URL \url{http://cdsweb.cern.ch/search.py?recid=352337}.
\newblock  (1964).

\bibitem{Zweig:1964jf}
G.~Zweig.
\newblock An {SU(3)} model for strong interaction symmetry and its
  breaking~{II}.
\newblock URL \url{http://cdsweb.cern.ch/search.py?recid=570209}.
\newblock  (1964).

\bibitem{Gell-Mann:1964nj}
M.~Gell-Mann, A schematic model of baryons and mesons, \emph{Phys. Lett.} {\bf
  8}, \penalty0 214--215,  (1964).
\newblock \doi{10.1016/S0031-9163(64)92001-3}.

\bibitem{Greenberg:1964pe}
O.~W. Greenberg, Spin and unitary spin independence in a paraquark model of
  baryons and mesons, \emph{Phys. Rev. Lett.} {\bf 13}, \penalty0 598--602,
  (1964).
\newblock \doi{10.1103/PhysRevLett.13.598}.

\bibitem{Han:1965pf}
M.-Y. Han and Y.~Nambu, Three triplet model with double {SU(3)} symmetry,
  \emph{Phys. Rev.} {\bf 139}, \penalty0 B1006--B1010,  (1965).
\newblock \doi{10.1103/PhysRev.139.B1006}.

\bibitem{Bjorken:1969ja}
J.~D. Bjorken and E.~A. Paschos, Inelastic electron-proton and $\gamma$-proton
  scattering and the structure of the nucleon, \emph{Phys. Rev.} {\bf 185},
  \penalty0 1975--1982,  (1969).
\newblock \doi{10.1103/PhysRev.185.1975}.

\bibitem{Fritzsch:1973pi}
H.~Fritzsch, M.~Gell-Mann, and H.~Leutwyler, Advantages of the color-octet
  gluon picture, \emph{Phys. Lett.} {\bf B47}, \penalty0 365--368,  (1973).
\newblock \doi{10.1016/0370-2693(73)90625-4}.

\bibitem{Yang:1954ek}
C.~N. Yang and R.~L. Mills, Conservation of isotopic spin and isotopic gauge
  invariance, \emph{Phys. Rev.} {\bf 96}, \penalty0 191--195,  (1954).
\newblock \doi{10.1103/PhysRev.96.191}.

\bibitem{RonShaw}
R.~Shaw.
\newblock \emph{The problem of particle types and other contributions to the
  theory of elementary particles}.
\newblock PhD thesis, Cambridge University,  (1955).

\bibitem{Politzer:1973fx}
H.~D. Politzer, Reliable perturbative results for strong interactions,
  \emph{Phys. Rev. Lett.} {\bf 30}, \penalty0 1346--1349,  (1973).
\newblock \doi{10.1103/PhysRevLett.30.1346}.

\bibitem{Gross:1973id}
D.~J. Gross and F.~Wilczek, Ultraviolet behavior of nonabelian gauge theories,
  \emph{Phys. Rev. Lett.} {\bf 30}, \penalty0 1343--1346,  (1973).
\newblock \doi{10.1103/PhysRevLett.30.1343}.

\bibitem{Wilson:1974sk}
K.~G. Wilson, Confinement of quarks, \emph{Phys. Rev.} {\bf D10}, \penalty0
  2445--2459,  (1974).
\newblock \doi{10.1103/PhysRevD.10.2445}.

\bibitem{GellMann:1954fq}
M.~Gell-Mann and F.~E. Low, Quantum electrodynamics at small distances,
  \emph{Phys. Rev.} {\bf 95}, \penalty0 1300--1312,  (1954).
\newblock \doi{10.1103/PhysRev.95.1300}.

\bibitem{Wilson:1969zs}
K.~G. Wilson, Non-{Lagrangian} models of current algebra, \emph{Phys. Rev.}
  {\bf 179}, \penalty0 1499--1512,  (1969).
\newblock \doi{10.1103/PhysRev.179.1499}.

\bibitem{Wilson:1970ag}
K.~G. Wilson, The renormalization group and strong interactions, \emph{Phys.
  Rev.} {\bf D3}, \penalty0 1818--1846,  (1971).
\newblock \doi{10.1103/PhysRevD.3.1818}.

\bibitem{Celmaster:1979km}
W.~Celmaster and R.~J. Gonsalves, The renormalization prescription dependence
  of the {QCD} coupling constant, \emph{Phys. Rev.} {\bf D20}, \penalty0 1420,
  (1979).
\newblock \doi{10.1103/PhysRevD.20.1420}.

\bibitem{Ellis:1991qj}
R.~K. Ellis, W.~J. Stirling, and B.~R. Webber, \emph{QCD and Collider Physics}.
  (Cambridge University, Cambridge, U.K., 1996).

\bibitem{Perl:2009zz}
M.~L. Perl, E.~R. Lee, and D.~Loomba, Searches for fractionally charged
  particles, \emph{Annu. Rev. Nucl. Part. Sci.} {\bf 59}, \penalty0 47--65,
  (2009).
\newblock \doi{10.1146/annurev-nucl-121908-122035}.

\bibitem{Wilson:2004de}
K.~G. Wilson, The origins of lattice gauge theory, \emph{Nucl. Phys. Proc.
  Suppl.} {\bf 140}, \penalty0 3--19,  (2005).
\newblock \doi{10.1016/j.nuclphysbps.2004.11.271}.

\bibitem{'tHooft:1971fh}
G.~'t~Hooft, Renormalization of massless {Yang-Mills} fields, \emph{Nucl.
  Phys.} {\bf B33}, \penalty0 173--199,  (1971).
\newblock \doi{10.1016/0550-3213(71)90395-6}.

\bibitem{'tHooft:1972fi}
G.~'t~Hooft and M.~J.~G. Veltman, Regularization and renormalization of gauge
  fields, \emph{Nucl. Phys.} {\bf B44}, \penalty0 189--213,  (1972).
\newblock \doi{10.1016/0550-3213(72)90279-9}.

\bibitem{Wilson:1971bg}
K.~G. Wilson, Renormalization group and critical phenomena~{I}: Renormalization
  group and the {Kadanoff} scaling picture, \emph{Phys. Rev.} {\bf B4},
  \penalty0 3174--3183,  (1971).
\newblock \doi{10.1103/PhysRevB.4.3174}.

\bibitem{Wilson:1971dh}
K.~G. Wilson, Renormalization group and critical phenomena~{II}: Phase space
  cell analysis of critical behavior, \emph{Phys. Rev.} {\bf B4}, \penalty0
  3184--3205,  (1971).
\newblock \doi{10.1103/PhysRevB.4.3184}.

\bibitem{Wilson:1974mb}
K.~G. Wilson, The renormalization group: Critical phenomena and the {Kondo}
  problem, \emph{Rev. Mod. Phys.} {\bf 47}, \penalty0 773--840,  (1975).
\newblock \doi{10.1103/RevModPhys.47.773}.

\bibitem{Wilson:1993dy}
K.~G. Wilson, The renormalization group and critical phenomena, \emph{Rev. Mod.
  Phys.} {\bf 55}, \penalty0 583--600,  (1983).
\newblock \doi{10.1103/RevModPhys.55.583}.

\bibitem{Christ:1982ck}
N.~H. Christ, R.~Friedberg, and T.~D. Lee, Gauge theory on a random lattice,
  \emph{Nucl. Phys.} {\bf B210}, \penalty0 310--336,  (1982).
\newblock \doi{10.1016/0550-3213(82)90123-7}.

\bibitem{Wentzel:1940gu}
G.~Wentzel, {Zum Problem des statischen Mesonfeldes}, \emph{Helv. Phys. Acta}.
  {\bf 13}, \penalty0 269--308,  (1940).
\newblock \doi{10.5169/seals-111066}.
\newblock (E) \textit{ibid.}\ \textbf{14}, 633 (1941).

\bibitem{Schiff:1953zza}
L.~I. Schiff, Lattice-space quantization of a nonlinear field theory,
  \emph{Phys. Rev.} {\bf 92}, \penalty0 766--779,  (1953).
\newblock \doi{10.1103/PhysRev.92.766}.

\bibitem{Wilson:1984vw}
K.~G. Wilson, \emph{Future directions in particle theory}, In eds. D.~G. Cassel
  and D.~L. Kreinick, \emph{Lepton-Photon Symposium 1983}, pp. 812--834.
\newblock Newman Laboratory, Cornell University, Ithaca, NY,  (1983).

\bibitem{Wegner:1971qt}
F.~J. Wegner, Duality in generalized {Ising} models and phase transitions
  without local order parameters, \emph{J. Math. Phys.} {\bf 12}, \penalty0
  2259--2272,  (1971).
\newblock \doi{10.1063/1.1665530}.

\bibitem{Smit:1972}
J.~Smit.
\newblock Unpublished.
\newblock Noted in Refs.~\onlinecite{Wilson:1984vw}
  and~\onlinecite{Smit:2002ug},  (1972).

\bibitem{Smit:2002ug}
J.~Smit, \emph{Introduction to quantum fields on a lattice: A robust mate}.
  (Cambridge Uni- versity, Cambridge, UK, 2002).
\newblock Endnote to Chapter 4 on p.~257.

\bibitem{Polyakov:1975rs}
A.~M. Polyakov, Compact gauge fields and the infrared catastrophe, \emph{Phys.
  Lett.} {\bf B59}, \penalty0 82--84,  (1975).
\newblock \doi{10.1016/0370-2693(75)90162-8}.
\newblock This publication builds on unpublished work on gauge fields on a
  lattice.

\bibitem{Wentzel:1949}
G.~Wentzel, \emph{Quantum Theory of Fields}. (Interscience, New York, 1949).

\bibitem{Schiff:1968}
L.~I. Schiff, \emph{Quantum Mechanics}. (McGraw-Hill, New York, 1968).

\bibitem{Heisenberg:1929wf}
W.~Heisenberg and W.~Pauli, {Quantendynamik der Wellenfelder}, \emph{Z. Phys.}
  {\bf 56}, \penalty0 1--61,  (1929).
\newblock \doi{10.1007/BF01340129}.

\bibitem{Rebbi:1983}
C.~Rebbi, \emph{Lattice Gauge Theories and Monte Carlo Simulations}. (World
  Scientific, Singapore, 1983).
\newblock A reprint volume.

\bibitem{Balian:1974ts}
R.~Balian, J.~M. Drouffe, and C.~Itzykson, Gauge fields on a lattice~1: General
  outlook, \emph{Phys. Rev.} {\bf D10}, \penalty0 3376--3395,  (1974).
\newblock \doi{10.1103/PhysRevD.10.3376}.

\bibitem{Balian:1974ir}
R.~Balian, J.~M. Drouffe, and C.~Itzykson, Gauge fields on a lattice~2: Gauge
  invariant {Ising} model, \emph{Phys. Rev.} {\bf D11}, \penalty0 2098--2103,
  (1975).
\newblock \doi{10.1103/PhysRevD.11.2098}.

\bibitem{Balian:1974xw}
R.~Balian, J.~M. Drouffe, and C.~Itzykson, Gauge fields on a lattice~3: Strong
  coupling expansions and transition points, \emph{Phys. Rev.} {\bf D11},
  \penalty0 2104--2119,  (1975).
\newblock \doi{10.1103/PhysRevD.11.2104}.
\newblock (E) \emph{ibid}.\ \textbf{D11}, 2514 (1975), doi:
  10.1103/PhysRevD.19.2514.

\bibitem{Haar:1933im}
A.~Haar, {Der Ma\ss begriff in der Theorie der kontinuierlichen Gruppen},
  \emph{Ann. Math.} {\bf 34}, \penalty0 147--169,  (1933).
\newblock URL \url{http://www.jstor.org/stable/1968346}.

\bibitem{Berezin:1966nc}
F.~A. Berezin, \emph{The Method of Second Quantization}. (Academic, New York,
  1966).

\bibitem{Wilson:1973jj}
K.~G. Wilson and J.~B. Kogut, The renormalization group and the $\epsilon$
  expansion, \emph{Phys. Rept.} {\bf 12}, \penalty0 75--200,  (1974).
\newblock \doi{10.1016/0370-1573(74)90023-4}.

\bibitem{Seiler:1982pw}
E.~Seiler, Gauge theories as a problem of constructive quantum field theory and
  statistical mechanics, \emph{Lect. Notes Phys.} {\bf 159}, \penalty0 1--192,
  (1982).

\bibitem{Osterwalder:1973dx}
K.~Osterwalder and R.~Schrader, Axioms for {Euclidean} {Green's} functions,
  \emph{Commun. Math. Phys.} {\bf 31}, \penalty0 83--112,  (1973).
\newblock \doi{10.1007/BF01645738}.

\bibitem{Osterwalder:1974tc}
K.~Osterwalder and R.~Schrader, Axioms for {Euclidean} {Green's}
  functions~{II}, \emph{Commun. Math. Phys.} {\bf 42}, \penalty0 281--305,
  (1975).
\newblock \doi{10.1007/BF01608978}.

\bibitem{Osterwalder:1977pc}
K.~Osterwalder and E.~Seiler, Gauge field theories on the lattice, \emph{Ann.
  Phys.} {\bf 110}, \penalty0 440--471,  (1978).
\newblock \doi{10.1016/0003-4916(78)90039-8}.

\bibitem{Summers:2012vr}
S.~J. Summers.
\newblock A perspective on constructive quantum field theory.
\newblock arXiv:1203.3991 [math-ph],  (2012).

\bibitem{Savit:1977fw}
R.~Savit, Topological excitations in {U(1)} invariant theories, \emph{Phys.
  Rev. Lett.} {\bf 39}, \penalty0 55--58,  (1977).
\newblock \doi{10.1103/PhysRevLett.39.55}.

\bibitem{Banks:1977cc}
T.~Banks, R.~Myerson, and J.~B. Kogut, Phase transitions in {Abelian} lattice
  gauge theories, \emph{Nucl. Phys.} {\bf B129}, \penalty0 493--510,  (1977).
\newblock \doi{10.1016/0550-3213(77)90129-8}.

\bibitem{Guth:1979gz}
A.~H. Guth, Existence proof of a nonconfining phase in four-dimensional {U(1)}
  lattice gauge theory, \emph{Phys. Rev.} {\bf D21}, \penalty0 2291--2307,
  (1980).
\newblock \doi{10.1103/PhysRevD.21.2291}.

\bibitem{Frohlich:1982gf}
J.~Fr\"ohlich and T.~Spencer, Massless phases and symmetry restoration in
  {Abelian} gauge theories and spin systems, \emph{Commun. Math. Phys.} {\bf
  83}, \penalty0 411--454,  (1982).
\newblock \doi{10.1007/BF01213610}.

\bibitem{Creutz:1980zw}
M.~Creutz, {Monte Carlo} study of quantized {SU(2)} gauge theory, \emph{Phys.
  Rev.} {\bf D21}, \penalty0 2308--2315,  (1980).
\newblock \doi{10.1103/PhysRevD.21.2308}.

\bibitem{Creutz:1980wj}
M.~Creutz, Asymptotic-freedom scales, \emph{Phys. Rev. Lett.} {\bf 45},
  \penalty0 313,  (1980).
\newblock \doi{10.1103/PhysRevLett.45.313}.

\bibitem{Kogut:1979vg}
J.~B. Kogut, R.~B. Pearson, and J.~Shigemitsu, Quantum-chromodynamic $\beta$
  function at intermediate and strong coupling, \emph{Phys. Rev. Lett.} {\bf
  43}, \penalty0 484--486,  (1979).
\newblock \doi{10.1103/PhysRevLett.43.484}.

\bibitem{Munster:1980iv}
G.~M\"unster, High temperature expansions for the free energy of vortices,
  respectively the string tension in lattice gauge theories, \emph{Nucl. Phys.}
  {\bf B180}, \penalty0 23,  (1981).
\newblock \doi{10.1016/0550-3213(81)90153-X}.

\bibitem{Bhanot:1981eb}
G.~Bhanot and M.~Creutz, Variant actions and phase structure in lattice gauge
  theory, \emph{Phys. Rev.} {\bf D24}, \penalty0 3212,  (1981).
\newblock \doi{10.1103/PhysRevD.24.3212}.

\bibitem{Bhanot:1981pj}
G.~Bhanot, {SU(3)} lattice gauge theory in four-dimensions with a modified
  {Wilson} action, \emph{Phys. Lett.} {\bf B108}, \penalty0 337,  (1982).
\newblock \doi{10.1016/0370-2693(82)91207-2}.

\bibitem{Munster:1981es}
G.~M\"unster, Strong coupling expansions for the mass gap in lattice gauge
  theories, \emph{Nucl. Phys.} {\bf B190}, \penalty0 439--453,  (1981).
\newblock \doi{10.1016/0550-3213(81)90570-8}.

\bibitem{Seo:1982jh}
K.~Seo, Glueball mass estimate by strong coupling expansion in lattice gauge
  theories, \emph{Nucl. Phys.} {\bf B209}, \penalty0 200--216,  (1982).
\newblock \doi{10.1016/0550-3213(82)90110-9}.

\bibitem{Smit:1982fx}
J.~Smit, Estimate of glueball masses from their strong coupling series in
  lattice {QCD}, \emph{Nucl. Phys.} {\bf B206}, \penalty0 309,  (1982).
\newblock \doi{10.1016/0550-3213(82)90537-5}.

\bibitem{Munster:1982kg}
G.~M\"unster, Physical strong coupling expansion parameters and glueball mass
  ratios, \emph{Phys. Lett.} {\bf B121}, \penalty0 53,  (1983).
\newblock \doi{10.1016/0370-2693(83)90201-0}.

\bibitem{Kogut:1976zr}
J.~B. Kogut, D.~K. Sinclair, and L.~Susskind, A quantitative approach to
  low-energy quantum chromodynamics, \emph{Nucl. Phys.} {\bf B114}, \penalty0
  199,  (1976).
\newblock \doi{10.1016/0550-3213(76)90586-1}.

\bibitem{Banks:1976ia}
T.~Banks et~al., Strong coupling calculations of the hadron spectrum of quantum
  chromodynamics, \emph{Phys. Rev.} {\bf D15}, \penalty0 1111,  (1977).
\newblock \doi{10.1103/PhysRevD.15.1111}.

\bibitem{Blairon:1980pk}
J.~M. Blairon, R.~Brout, F.~Englert, and J.~Greensite, Chiral symmetry breaking
  in the action formulation of lattice gauge theory, \emph{Nucl. Phys.} {\bf
  B180}, \penalty0 439,  (1981).
\newblock \doi{10.1016/0550-3213(81)90061-4}.

\bibitem{Kogut:1982ds}
J.~B. Kogut, A review of the lattice gauge theory approach to quantum
  chromo\-dynamics, \emph{Rev. Mod. Phys.} {\bf 55}, \penalty0 775--836,
  (1983).
\newblock \doi{10.1103/RevModPhys.55.775}.

\bibitem{KlubergStern:1982bs}
H.~Kluberg-Stern, A.~Morel, and B.~Petersson, Spectrum of lattice gauge
  theories with fermions from a $1/d$ expansion at strong coupling, \emph{Nucl.
  Phys.} {\bf B215}, \penalty0 527,  (1983).
\newblock \doi{10.1016/0550-3213(83)90259-6}.

\bibitem{Martin:1983hw}
O.~Martin, Mesons and baryons at large {$N$} and strong coupling, \emph{Phys.
  Lett.} {\bf B130}, \penalty0 411,  (1983).
\newblock \doi{10.1016/0370-2693(83)91533-2}.

\bibitem{Karsten:1980wd}
L.~H. Karsten and J.~Smit, Lattice fermions: Species doubling, chiral
  invariance, and the triangle anomaly, \emph{Nucl. Phys.} {\bf B183},
  \penalty0 103,  (1981).
\newblock \doi{10.1016/0550-3213(81)90549-6}.

\bibitem{Kogut:1974ag}
J.~B. Kogut and L.~Susskind, {Hamiltonian} formulation of {Wilson}'s lattice
  gauge theories, \emph{Phys. Rev.} {\bf D11}, \penalty0 395--408,  (1975).
\newblock \doi{10.1103/PhysRevD.11.395}.

\bibitem{Susskind:1976jm}
L.~Susskind, Lattice fermions, \emph{Phys. Rev.} {\bf D16}, \penalty0
  3031--3039,  (1977).
\newblock \doi{10.1103/PhysRevD.16.3031}.

\bibitem{Kawamoto:1981hw}
N.~Kawamoto and J.~Smit, Effective {Lagrangian} and dynamical symmetry breaking
  in strongly coupled lattice {QCD}, \emph{Nucl. Phys.} {\bf B192}, \penalty0
  100,  (1981).
\newblock \doi{10.1016/0550-3213(81)90196-6}.

\bibitem{Sharatchandra:1981si}
H.~S. Sharatchandra, H.~J. Thun, and P.~Weisz, Susskind fermions on a
  {Euclidean} lattice, \emph{Nucl. Phys.} {\bf B192}, \penalty0 205,  (1981).
\newblock \doi{10.1016/0550-3213(81)90200-5}.

\bibitem{Smit:1987zh}
J.~Smit and J.~C. Vink, Renormalized {Ward}-{Takahashi} relations and
  topological susceptibility with staggered fermions, \emph{Nucl. Phys.} {\bf
  B298}, \penalty0 557,  (1988).
\newblock \doi{10.1016/0550-3213(88)90354-9}.

\bibitem{Wilson:1977nj}
K.~G. Wilson.
\newblock Quantum chromodynamics on a lattice.
\newblock In ed. A.~Zichichi, \emph{New Phenomena in Subnuclear Physics}.
  Plenum, New York,  (1977).

\bibitem{Sheikholeslami:1985ij}
B.~Sheikholeslami and R.~Wohlert, Improved continuum limit lattice action for
  {QCD} with {Wilson} fermions, \emph{Nucl. Phys.} {\bf B259}, \penalty0
  572--596,  (1985).
\newblock \doi{10.1016/0550-3213(85)90002-1}.

\bibitem{Jansen:1995ck}
K.~Jansen et~al., Nonperturbative renormalization of lattice {QCD} at all
  scales, \emph{Phys. Lett.} {\bf B372}, \penalty0 275--282,  (1996).
\newblock \doi{10.1016/0370-2693(96)00075-5}.

\bibitem{Luscher:1996sc}
M.~L\"uscher, S.~Sint, R.~Sommer, and P.~Weisz, Chiral symmetry and {O($a$)}
  improvement in lattice {QCD}, \emph{Nucl. Phys.} {\bf B478}, \penalty0
  365--400,  (1996).
\newblock \doi{10.1016/0550-3213(96)00378-1}.

\bibitem{Frezzotti:2000nk}
R.~Frezzotti, P.~A. Grassi, S.~Sint, and P.~Weisz, Lattice {QCD} with a
  chirally twisted mass term, \emph{{JHEP}}. {\bf 0108}, \penalty0 058,
  (2001).

\bibitem{Frezzotti:2003ni}
R.~Frezzotti and G.~C. Rossi, Chirally improving {Wilson} fermions~1: {O($a$)}
  improvement, \emph{{JHEP}}. {\bf 0408}, \penalty0 007,  (2004).
\newblock \doi{10.1088/1126-6708/2004/08/007}.

\bibitem{Nielsen:1980rz}
H.~B. Nielsen and M.~Ninomiya, Absence of neutrinos on a lattice~1: Proof by
  homotopy theory, \emph{Nucl. Phys.} {\bf B185}, \penalty0 20--40,  (1981).
\newblock \doi{10.1016/0550-3213(81)90361-8, 10.1016/0550-3213(82)90011-6}.
\newblock {(E)} \textit{Nucl. Phys.} \textbf{B195}, 541--542, (1982).

\bibitem{Friedan:1982nk}
D.~Friedan, A proof of the {Nielsen-Ninomiya} theorem, \emph{Commun. Math.
  Phys.} {\bf 85}, \penalty0 481--490,  (1982).
\newblock \doi{10.1007/BF01403500}.

\bibitem{Ginsparg:1981bj}
P.~H. Ginsparg and K.~G. Wilson, A remnant of chiral symmetry on the lattice,
  \emph{Phys. Rev.} {\bf D25}, \penalty0 2649,  (1982).
\newblock \doi{10.1103/PhysRevD.25.2649}.

\bibitem{Shamir:1993zy}
Y.~Shamir, Chiral fermions from lattice boundaries, \emph{Nucl. Phys.} {\bf
  B406}, \penalty0 90--106,  (1993).
\newblock \doi{10.1016/0550-3213(93)90162-I}.

\bibitem{Neuberger:1997fp}
H.~Neuberger, Exactly massless quarks on the lattice, \emph{Phys. Lett.} {\bf
  B417}, \penalty0 141--144,  (1998).
\newblock \doi{10.1016/S0370-2693(97)01368-3}.

\bibitem{Hasenfratz:1998jp}
P.~Hasenfratz, Lattice {QCD} without tuning, mixing and current
  renormalization, \emph{Nucl. Phys.} {\bf B525}, \penalty0 401--409,  (1998).
\newblock \doi{10.1016/S0550-3213(98)00399-X}.

\bibitem{Luscher:1998pqa}
M.~L\"uscher, Exact chiral symmetry on the lattice and the {Ginsparg-Wilson}
  relation, \emph{Phys. Lett.} {\bf B428}, \penalty0 342--345,  (1998).
\newblock \doi{10.1016/S0370-2693(98)00423-7}.

\bibitem{Chandrasekharan:1998wg}
S.~Chandrasekharan, Lattice {QCD} with {Ginsparg-Wilson} fermions, \emph{Phys.
  Rev.} {\bf D60}, \penalty0 074503,  (1999).
\newblock \doi{10.1103/PhysRevD.60.074503}.

\bibitem{Kaplan:1992bt}
D.~B. Kaplan, A method for simulating chiral fermions on the lattice,
  \emph{Phys. Lett.} {\bf B288}, \penalty0 342--347,  (1992).
\newblock \doi{10.1016/0370-2693(92)91112-M}.

\bibitem{Narayanan:1994gw}
R.~Narayanan and H.~Neuberger, A construction of lattice chiral gauge theories,
  \emph{Nucl. Phys.} {\bf B443}, \penalty0 305--385,  (1995).
\newblock \doi{10.1016/0550-3213(95)00111-5}.

\bibitem{Luscher:1998du}
M.~L\"uscher, Abelian chiral gauge theories on the lattice with exact gauge
  invariance, \emph{Nucl. Phys.} {\bf B549}, \penalty0 295--334,  (1999).
\newblock \doi{10.1016/S0550-3213(99)00115-7}.

\bibitem{Luscher:1999un}
M.~L\"uscher, Weyl fermions on the lattice and the non-{Abelian} gauge anomaly,
  \emph{Nucl. Phys.} {\bf B568}, \penalty0 162--179,  (2000).
\newblock \doi{10.1016/S0550-3213(99)00731-2}.

\bibitem{Creutz:1976ch}
M.~Creutz, Gauge fixing, the transfer matrix, and confinement on a lattice,
  \emph{Phys. Rev.} {\bf D15}, \penalty0 1128,  (1977).
\newblock \doi{10.1103/PhysRevD.15.1128}.

\bibitem{Luscher:1976ms}
M.~L\"uscher, Construction of a self-adjoint, strictly positive transfer matrix
  for {Euclidean} lattice gauge theories, \emph{Commun. Math. Phys.} {\bf 54},
  \penalty0 283,  (1977).
\newblock \doi{10.1007/BF01614090}.

\bibitem{Luscher:1984is}
M.~L\"uscher and P.~Weisz, Definition and general properties of the transfer
  matrix in continuum limit improved lattice gauge theories, \emph{Nucl. Phys.}
  {\bf B240}, \penalty0 349,  (1984).
\newblock \doi{10.1016/0550-3213(84)90270-0}.

\bibitem{DiPierro:2005vz}
M.~{Di Pierro}, An algorithmic approach to quantum field theory, \emph{Int. J.
  Mod. Phys.} {\bf A21}, \penalty0 405--448,  (2006).
\newblock \doi{10.1142/S0217751X06028965}.

\bibitem{Weingarten:1981jy}
D.~Weingarten, {Monte Carlo} evaluation of hadron masses in lattice gauge
  theories with fermions, \emph{Phys. Lett.} {\bf B109}, \penalty0 57--62,
  (1982).
\newblock \doi{10.1016/0370-2693(82)90463-4}.

\bibitem{Marinari:1981qf}
E.~Marinari, G.~Parisi, and C.~Rebbi, {Monte Carlo} simulation of the massive
  {Schwinger} model, \emph{Nucl. Phys.} {\bf B190}, \penalty0 734--750,
  (1981).
\newblock \doi{10.1016/0550-3213(81)90048-1}.

\bibitem{Davies:2003ik}
C.~T.~H. Davies et~al., High-precision lattice {QCD} confronts experiment,
  \emph{Phys. Rev. Lett.} {\bf 92}, \penalty0 022001,  (2004).
\newblock \doi{10.1103/PhysRevLett.92.022001}.

\bibitem{Hagiwara:2002fs}
K.~Hagiwara et~al., Review of particle physics, \emph{Phys. Rev.} {\bf D66},
  \penalty0 010001,  (2002).
\newblock \doi{10.1103/PhysRevD.66.010001}.

\bibitem{Aubin:2004ej}
C.~Aubin et~al., Semileptonic decays of {$D$} mesons in three-flavor lattice
  {QCD}, \emph{Phys. Rev. Lett.} {\bf 94}, \penalty0 011601,  (2005).
\newblock \doi{10.1103/PhysRevLett.94.011601}.

\bibitem{Allison:2004be}
I.~F. Allison et~al., Mass of the {$B_c$} meson in three-flavor lattice {QCD},
  \emph{Phys. Rev. Lett.} {\bf 94}, \penalty0 172001,  (2005).
\newblock \doi{10.1103/PhysRevLett.94.172001}.

\bibitem{Aubin:2005ar}
C.~Aubin et~al., Charmed meson decay constants in three-flavor lattice {QCD},
  \emph{Phys. Rev. Lett.} {\bf 95}, \penalty0 122002,  (2005).
\newblock \doi{10.1103/PhysRevLett.95.122002}.

\bibitem{Gray:2005ur}
A.~Gray et~al., The {$\Upsilon$} spectrum and $m_b$ from full lattice {QCD},
  \emph{Phys. Rev.} {\bf D72}, \penalty0 094507,  (2005).
\newblock \doi{10.1103/PhysRevD.72.094507}.

\bibitem{Besson:2009uv}
D.~Besson et~al., Improved measurements of {$D$} meson semileptonic decays to
  $\pi$ and {$K$} mesons, \emph{Phys. Rev.} {\bf D80}, \penalty0 032005,
  (2009).
\newblock \doi{10.1103/PhysRevD.80.032005}.

\bibitem{Bernard:2009ke}
C.~Bernard et~al., Visualization of semileptonic form factors from lattice
  {QCD}, \emph{Phys. Rev.} {\bf D80}, \penalty0 034026,  (2009).
\newblock \doi{10.1103/PhysRevD.80.034026}.

\bibitem{Link:2004dh}
J.~M. Link et~al., Measurements of the $q^{2}$ dependence of the {$D^0\to
  K^{-}\mu^{+}\nu$} and {$D^0\to\pi^{-}\mu^{+}\nu$} form factors, \emph{Phys.
  Lett.} {\bf B607}, \penalty0 233--242,  (2005).
\newblock \doi{10.1016/j.physletb.2004.12.036}.

\bibitem{Widhalm:2006wz}
L.~Widhalm et~al., Measurement of {$D^0\to\pi l\nu\,(Kl\nu)$} form factors and
  absolute branching fractions, \emph{Phys. Rev. Lett.} {\bf 97}, \penalty0
  061804,  (2006).
\newblock \doi{10.1103/PhysRevLett.97.061804}.

\bibitem{Aubert:2007wg}
B.~Aubert et~al., Measurement of the hadronic form-factor in {$D^0 \to K^{-}
  e^{+} \nu_{e}$}, \emph{Phys. Rev.} {\bf D76}, \penalty0 052005,  (2007).
\newblock \doi{doi:10.1103/PhysRevD.76.052005}.

\bibitem{Dobbs:2007aa}
S.~Dobbs et~al., A study of the semileptonic charm decays
  {$D^0\to\pi^-e^+\nu_e$}, {$D^+\to\pi^0e^+\nu_e$}, {$D^0\to K^-e^+\nu_e$}, and
  {$D^+\to\bar{K}^0e^+\nu_e$}, \emph{Phys. Rev.} {\bf D77}, \penalty0 112005,
  (2008).
\newblock \doi{10.1103/PhysRevD.77.112005}.

\bibitem{Ge:2008aa}
J.~Y. Ge et~al., Study of {$D^0\to\pi^-e^+\nu_e$}, {$D^+\to\pi^0e^+\nu_e$},
  {$D^0\to K^-e^+\nu_e$}, and {$D^+\to\bar{K}^0e^+\nu_e$} in tagged decays of
  the $\psi(3770)$ resonance, \emph{Phys. Rev.} {\bf D79}, \penalty0 052010,
  (2009).
\newblock \doi{10.1103/PhysRevD.79.052010}.

\bibitem{Kronfeld:2012uk}
A.~S. Kronfeld, Twenty-first century lattice gauge theory: Results from the
  {QCD} {Lagrangian}, \emph{Annu. Rev. Nucl. Part. Sci.} {\bf 62}, \penalty0
  265--284,  (2012).
\newblock \doi{10.1146/annurev-nucl-102711-094942}.

\bibitem{Aubin:2004wf}
C.~Aubin et~al., Light hadrons with improved staggered quarks: Approaching the
  continuum limit, \emph{Phys. Rev.} {\bf D70}, \penalty0 094505,  (2004).
\newblock \doi{10.1103/PhysRevD.70.094505}.

\bibitem{Bazavov:2009bb}
A.~Bazavov et~al., Nonperturbative {QCD} simulations with 2+1 flavors of
  improved staggered quarks, \emph{Rev. Mod. Phys.} {\bf 82}, \penalty0
  1349--1417,  (2010).
\newblock \doi{10.1103/RevModPhys.82.1349}.

\bibitem{Aoki:2008sm}
S.~Aoki et~al., 2+1 flavor lattice {QCD} toward the physical point, \emph{Phys.
  Rev.} {\bf D79}, \penalty0 034503,  (2009).
\newblock \doi{10.1103/PhysRevD.79.034503}.

\bibitem{Durr:2008zz}
S.~D\"urr et~al., Ab-initio determination of light hadron masses,
  \emph{Science}. {\bf 322}, \penalty0 1224--1227,  (2008).
\newblock \doi{10.1126/science.1163233}.

\bibitem{Bietenholz:2011qq}
W.~Bietenholz et~al., Flavour blindness and patterns of flavour symmetry
  breaking in lattice simulations of up, down and strange quarks, \emph{Phys.
  Rev.} {\bf D84}, \penalty0 054509,  (2011).
\newblock \doi{10.1103/PhysRevD.84.054509}.

\bibitem{Christ:2010dd}
N.~H. Christ et~al., The $\eta$ and $\eta^\prime$ mesons from lattice {QCD},
  \emph{Phys. Rev. Lett.} {\bf 105}, \penalty0 241601,  (2010).
\newblock \doi{10.1103/PhysRevLett.105.241601}.

\bibitem{Dudek:2011tt}
J.~J. Dudek et~al., Isoscalar meson spectroscopy from lattice {QCD},
  \emph{Phys. Rev.} {\bf D83}, \penalty0 111502,  (2011).
\newblock \doi{10.1103/PhysRevD.83.111502}.

\bibitem{Gregory:2011sg}
E.~B. Gregory, A.~C. Irving, C.~M. Richards, and C.~McNeile, A study of the
  $\eta$ and $\eta'$ mesons with improved staggered fermions, \emph{Phys. Rev.}
  {\bf D86}, \penalty0 014504,  (2012).
\newblock \doi{10.1103/PhysRevD.86.014504}.

\bibitem{Bernard:2010fr}
C.~Bernard et~al., Tuning {Fermilab} heavy quarks in 2+1 flavor lattice {QCD}
  with application to hyperfine splittings, \emph{Phys. Rev.} {\bf D83},
  \penalty0 034503,  (2011).
\newblock \doi{10.1103/PhysRevD.83.034503}.

\bibitem{Gregory:2010gm}
E.~B. Gregory et~al., Precise {$B$, $B_s$, and $B_c$} meson spectroscopy from
  full lattice {QCD}, \emph{Phys. Rev.} {\bf D83}, \penalty0 014506,  (2011).
\newblock \doi{10.1103/PhysRevD.83.014506}.

\bibitem{Mohler:2011ke}
D.~Mohler and R.~M. Woloshyn, {$D$ and $D_s$} meson spectroscopy, \emph{Phys.
  Rev.} {\bf D84}, \penalty0 054505,  (2011).
\newblock \doi{10.1103/PhysRevD.84.054505}.

\bibitem{Fodor:2012gf}
Z.~Fodor and C.~Hoelbling, Light hadron masses from lattice {QCD}, \emph{Rev.
  Mod. Phys.} {\bf 84}, \penalty0 449--495,  (2012).
\newblock \doi{10.1103/RevModPhys.84.449}.

\bibitem{Beringer:1900zz}
J.~Beringer et~al., Review of particle physics, \emph{Phys. Rev.} {\bf D86},
  \penalty0 010001,  (2012).
\newblock \doi{10.1103/PhysRevD.86.010001}.

\bibitem{Morningstar:1999rf}
C.~J. Morningstar and M.~J. Peardon, The glueball spectrum from an anisotropic
  lattice study, \emph{Phys. Rev.} {\bf D60}, \penalty0 034509,  (1999).
\newblock \doi{10.1103/PhysRevD.60.034509}.

\bibitem{Richards:2010ck}
C.~M. Richards, A.~C. Irving, E.~B. Gregory, and C.~McNeile, Glueball mass
  measurements from improved staggered fermion simulations, \emph{Phys. Rev.}
  {\bf D82}, \penalty0 034501,  (2010).
\newblock \doi{10.1103/PhysRevD.82.034501}.

\bibitem{Bulava:2009jb}
J.~M. Bulava et~al., Excited state nucleon spectrum with two flavors of
  dynamical fermions, \emph{Phys. Rev.} {\bf D79}, \penalty0 034505,  (2009).
\newblock \doi{10.1103/PhysRevD.79.034505}.

\bibitem{Edwards:2011jj}
R.~G. Edwards, J.~J. Dudek, D.~G. Richards, and S.~J. Wallace, Excited state
  baryon spectroscopy from lattice {QCD}, \emph{Phys. Rev.} {\bf D84},
  \penalty0 074508,  (2011).
\newblock \doi{10.1103/PhysRevD.84.074508}.

\bibitem{Burkert:2009zf}
V.~D. Burkert, The {$N^*$} physics program at {Jefferson Lab}, \emph{Chin.
  Phys.} {\bf C33}, \penalty0 1043--1050,  (2009).
\newblock \doi{10.1088/1674-1137/33/12/001}.

\bibitem{Juge:2002br}
K.~J. Juge, J.~Kuti, and C.~Morningstar, Fine structure of the {QCD} string
  spectrum, \emph{Phys. Rev. Lett.} {\bf 90}, \penalty0 161601,  (2003).
\newblock \doi{10.1103/PhysRevLett.90.161601}.

\bibitem{Juge:2004xr}
K.~J. Juge, J.~Kuti, and C.~Morningstar.
\newblock {QCD} string formation and the {Casimir} energy.
\newblock In eds. H.~Suganuma et~al., \emph{Confinement 2003}, pp. 233--248,
  Singapore,  (2004). World Scientific.

\bibitem{Luscher:1980iy}
M.~L\"uscher, G.~M\"unster, and P.~Weisz, How thick are chromoelectric flux
  tubes?, \emph{Nucl. Phys.} {\bf B180}, \penalty0 1,  (1981).
\newblock \doi{10.1016/0550-3213(81)90151-6}.

\bibitem{Bali:2005fu}
G.~S. Bali, H.~Neff, T.~D\"ussel, T.~Lippert, and K.~Schilling, Observation of
  string breaking in {QCD}, \emph{Phys. Rev.} {\bf D71}, \penalty0 114513,
  (2005).
\newblock \doi{10.1103/PhysRevD.71.114513}.

\bibitem{Greensite:2003bk}
J.~Greensite, The confinement problem in lattice gauge theory, \emph{Prog.
  Part. Nucl. Phys.} {\bf 51}, \penalty0 1--83,  (2003).
\newblock \doi{10.1016/S0146-6410(03)90012-3}.

\bibitem{Duncan:1992eb}
A.~Duncan, E.~Eichten, and H.~Thacker, Lattice {QCD}, the quark model, and
  heavy-light wave functions, \emph{Phys. Lett.} {\bf B303}, \penalty0
  109--112,  (1993).
\newblock \doi{10.1016/0370-2693(93)90052-J}.

\bibitem{Nambu:1960xd}
Y.~Nambu, Axial vector current conservation in weak interactions, \emph{Phys.
  Rev. Lett.} {\bf 4}, \penalty0 380--382,  (1960).
\newblock \doi{10.1103/PhysRevLett.4.380}.

\bibitem{Goldstone:1961eq}
J.~Goldstone, Field theories with superconductor solutions, \emph{Nuovo Cim.}
  {\bf 19}, \penalty0 154--164,  (1961).
\newblock \doi{10.1007/BF02812722}.

\bibitem{Fukaya:2010na}
H.~Fukaya et~al., Determination of the chiral condensate from {QCD} {Dirac}
  spectrum on the lattice, \emph{Phys. Rev.} {\bf D83}, \penalty0 074501,
  (2011).
\newblock \doi{10.1103/PhysRevD.83.074501}.

\bibitem{Ishii:2006ec}
N.~Ishii, S.~Aoki, and T.~Hatsuda, The nuclear force from lattice {QCD},
  \emph{Phys. Rev. Lett.} {\bf 99}, \penalty0 022001,  (2007).
\newblock \doi{10.1103/PhysRevLett.99.022001}.

\bibitem{Beane:2012vq}
S.~R. Beane et~al., Light nuclei and hypernuclei from quantum chromodynamics in
  the limit of {SU(3)} flavor symmetry.  (2012).
\newblock arXiv:1206.5219 [hep-lat].

\bibitem{Carazza:1995lt}
B.~Carazza and H.~Kragh, Heisenberg's lattice world: The 1930 theory sketch,
  \emph{Am. J. Phys.} {\bf 63}, \penalty0 595--605,  (1995).
\newblock \doi{10.1119/1.17848}.

\bibitem{Heisenberg:1930se}
W.~Heisenberg, {Die Selbstenergie des Elektrons}, \emph{Z. Phys.} {\bf 65},
  \penalty0 4--13,  (1930).
\newblock \doi{10.1007/BF01397404}.

\bibitem{Kragh:1995am}
H.~Kragh, ~~{Arthur March}, {Werner Heisenberg}, and the search for a smallest
  length, \emph{Revue d'Histoire des Sciences}. {\bf 48}\penalty0 (4),
  \penalty0 401--434,  (1995).
\newblock \doi{10.3406/rhs.1995.1239}.

\bibitem{Hill}
C.~T. Hill.
\newblock private communication.

\bibitem{Soma}
O.~P. Pedersen and T.~Bundgaard.
\newblock The birth of {SOMA}?
\newblock URL \url{http://www.fam-bundgaard.dk/SOMA/NEWS/N030310.HTM}.
\newblock  (Retrieved August 2012).

\bibitem{Gardner:1961}
M.~Gardner, \emph{The Second Scientific American Book of Mathematical Puzzles
  and Diversions}. (University of Chicago, Chicago, 1987).

\end{thebibliography}

\end{document}